\documentclass[sigconf]{acmart}

\usepackage{amsmath}
\usepackage{amsfonts}

\usepackage{dsfont} 

\usepackage{mathtools} 

\usepackage{enumitem}

\usepackage{amsthm}
\newtheorem{theorem}{Theorem}

\newtheorem{definition}{Definition}

\usepackage{hyperref}

\usepackage{multirow}
\usepackage{booktabs}
\usepackage{graphicx}
\usepackage{caption}
\usepackage{subcaption}
\usepackage[outdir=./]{epstopdf}
\usepackage{siunitx}
\usepackage[ruled,vlined]{algorithm2e}
\usepackage{colortbl}
\usepackage{tabulary}
\usepackage{etoolbox}

\usepackage{booktabs}
\usepackage{makecell}

\usepackage{placeins} 
\usepackage{dblfloatfix} 

\newcommand{\timeValue}{t} 

\newcommand{\powerInequality}{\mathcal{I}} 
\newcommand{\calBlue}{\mathcal{B}} 
\newcommand{\calRed}{\mathcal{R}} 
\newcommand{\graph}{G} 
\newcommand{\nodeSet}{V} 
\newcommand{\edgeSet}{E} 

\newcommand{\redBirthProb}{r} 

\newcommand{\probEventOne }{p} 
\newcommand{\probEventTwo }{q} 

\newcommand{\deltaboth}{\delta} 
\newcommand{\deltaIn}{\deltaboth_{i}} 
\newcommand{\deltaOut}{\deltaboth_{o}} 

\newcommand{\homophilyRedAnyEvent}{\rho_{\calRed}} 
\newcommand{\homophilyBlueAnyEvent}{\rho_{\calBlue}}  

\newcommand{\homophilyRedEventOne}{\homophilyRedAnyEvent^{(1)}} 
\newcommand{\homophilyBlueEventOne}{\homophilyBlueAnyEvent^{(1)}}  

\newcommand{\homophilyRedEventTwo}{\homophilyRedAnyEvent^{(2)}} 
\newcommand{\homophilyBlueEventTwo}{\homophilyBlueAnyEvent^{(2)}}  

\newcommand{\homophilyRedEventThree}{\homophilyRedAnyEvent^{(3)}} 
\newcommand{\homophilyBlueEventThree}{\homophilyBlueAnyEvent^{(3)}}  

\newcommand{\matrixAnyEvent}{E} 
\newcommand{\matrixEventOne}{\matrixAnyEvent_{1}} 
\newcommand{\matrixEventTwo}{\matrixAnyEvent_{2}} 
\newcommand{\matrixEventThree}{\matrixAnyEvent_{3}} 

\newcommand{\inDegree}{d_{i}} 
\newcommand{\outDegree}{d_{o}} 
\newcommand{\degree}{d} 

\newcommand{\thetaIn}{\theta^{(i)}} 
\newcommand{\thetaOut}{\theta^{(o)}} 
\newcommand{\thetaVec}{\theta} 

\newcommand{\nonLinFunction}{F} 

\newcommand{\groupSize}{n} 

\newcommand{\poneBB}{{p}^{(1)}_{\calBlue_{\mathrm{new}}\rightarrow\calBlue_{\mathrm{old}}}} 
\newcommand{\poneBR}{{p}^{(1)}_{\calBlue_{\mathrm{new}}\rightarrow\calRed_{\mathrm{old}}}} 
\newcommand{\poneRR}{{p}^{(1)}_{\calRed_{\mathrm{new}}\rightarrow\calRed_{\mathrm{old}}}} 
\newcommand{\poneRB}{{p}^{(1)}_{\calRed_{\mathrm{new}}\rightarrow\calBlue_{\mathrm{old}}}} 

\newcommand{\ptwoBB}{{p}^{(2)}_{\calBlue_{\mathrm{new}}\leftarrow\calBlue_{\mathrm{old}}}} 
\newcommand{\ptwoBR}{{p}^{(2)}_{\calBlue_{\mathrm{new}}\leftarrow\calRed_{\mathrm{old}}}} 
\newcommand{\ptwoRR}{{p}^{(2)}_{\calRed_{\mathrm{new}}\leftarrow\calRed_{\mathrm{old}}}} 
\newcommand{\ptwoRB}{{p}^{(2)}_{\calRed_{\mathrm{new}}\leftarrow\calBlue_{\mathrm{old}}}} 

\newcommand{\pthreeBB}{{p}^{(3)}_{\calBlue_{\mathrm{old}}\rightarrow \calBlue_{\mathrm{old}}}} 
\newcommand{\pthreeBR}{{p}^{(3)}_{\calBlue_{\mathrm{old}}\rightarrow \calRed_{\mathrm{old}}}} 
\newcommand{\pthreeRR}{{p}^{(3)}_{\calRed_{\mathrm{old}}\rightarrow \calRed_{\mathrm{old}}}} 
\newcommand{\pthreeRB}{{p}^{(3)}_{\calRed_{\mathrm{old}}\rightarrow\calBlue_{\mathrm{old}}}} 

\newcommand{\barponeBR}{\bar{p}^{(1)}_{\calBlue_{\mathrm{new}}\rightarrow\calRed_{\mathrm{old}}}} 
\newcommand{\barponeRR}{\bar{p}^{(1)}_{\calRed_{\mathrm{new}}\rightarrow\calRed_{\mathrm{old}}}}

\newcommand{\barptwoBR}{\bar{p}^{(2)}_{\calBlue_{\mathrm{new}}\leftarrow\calRed_{\mathrm{old}}}} 
\newcommand{\barptwoRR}{\bar{p}^{(2)}_{\calRed_{\mathrm{new}}\leftarrow\calRed_{\mathrm{old}}}}

\newcommand{\barpthreeBR}{\bar{p}^{(3)}_{\calBlue_{\mathrm{old}}\rightarrow \calRed_{\mathrm{old}}}} 
\newcommand{\barpthreeRR}{\bar{p}^{(3)}_{\calRed_{\mathrm{old}}\rightarrow \calRed_{\mathrm{old}}}} 
\newcommand{\barpthreeRB}{\bar{p}^{(3)}_{\calRed_{\mathrm{old}}\rightarrow\calBlue_{\mathrm{old}}}} 

\AtBeginDocument{%
  }

\setcopyright{acmlicensed}
\copyrightyear{2026}
\acmYear{2026}
\acmDOI{}
\acmConference[WWW '26]{the ACM Web Conference 2026}{April 13--17,
  2026}{Dubai, UAE}
\acmISBN{}

\begin{document}

\title{Emergence of Structural Disparities in the Web of Scientific Citations}

\author{Buddhika Nettasinghe}
\orcid{0000-0002-6070-892X}
\affiliation{%
  \institution{University of Iowa}
  \city{Iowa City}
  \state{Iowa}
  \country{USA}
}
\email{buddhika-nettasinghe@uiowa.edu}

\author{Nazanin Alipourfard}
\orcid{0000-0001-9971-2179}
\affiliation{%
  \institution{University of Southern California}
  \city{Los Angeles}
\state{California}
  \country{USA}}
\email{nazanin.alipourfard@gmail.com}

\author{Vikram Krishnamurthy}
\orcid{0000-0002-4170-6056}
\affiliation{%
  \institution{Cornell University}
  \city{Ithaca}
  \state{New York}
  \country{USA}
}
\email{vikramk@cornell.edu}

\author{Kristina Lerman}
\orcid{0000-0002-5071-0575}
\affiliation{%
  \institution{Indiana University}
  \city{Bloomington}
\state{Indiana}
  \country{USA}}
\email{krlerman@iu.edu}

\renewcommand{\shortauthors}{Nettasinghe et al.}

\begin{abstract}
Scientific attention is unevenly distributed, creating inequities in recognition and distorting access to opportunities.
Using citations as a proxy, we quantify disparities in attention by gender and institutional prestige. We find that women receive systematically fewer citations than men, and that attention is increasingly concentrated among authors from elite institutions---patterns not fully explained by underrepresentation alone. 
To explain these dynamics, we introduce a model of citation network growth that incorporates homophily (tendency to cite similar authors), preferential attachment (favoring highly cited authors) and group size (underrepresentation).  The model shows that disparities arise not only from group size imbalances but also from cumulative advantage amplifying biased citation preferences. Importantly, increasing representation alone is often insufficient to reduce disparities. Effective strategies should also include reducing homophily, amplifying the visibility of underrepresented groups, and supporting equitable integration of newcomers.
Our findings highlight the challenges of mitigating inequities in asymmetric networks like citations, where recognition flows in one direction. By making visible the mechanisms through which attention is distributed, we contribute to efforts toward a more responsible web of science that is fairer, more transparent, and more inclusive, and that better sustains innovation and knowledge production.
\end{abstract}



\maketitle

\vspace{-0.1cm}
\emph{\bf This pre-print is a more detailed version of the above paper accepted to
WWW’26, containing additional supplementary results and details.}

\section{Introduction}
Scientific attention is distributed unevenly~\cite{merton1968matthew,allison1982cumulative,nielsen2021global}, resulting in unequal access to opportunities~\cite{bol2018matthew}, hiring decisions~\cite{clauset2015systematic}, and scientific recognition~\cite{lerman2022gendered}. 
A well-documented driver of this inequality is cumulative advantage (aka the ``Matthew Effect'')~\cite{merton1968matthew}, whereby early recognition amplifies future visibility and disproportionately channels attention to already advantaged scientists~\cite{merton1968matthew,allison1982cumulative}. Framed as meritocratic stratification~\cite{xie2014undemocracy}, cumulative advantage grants select few researchers, often those at prestigious institutions, greater access to resources~\cite{burghardt2020heterogeneous}, collaborators~\cite{nielsen2021global}, and funding~\cite{bol2018matthew}.

A parallel line of research has documented systematic gender disparities in citations across disciplines, with women consistently receiving fewer citations, a proxy of scientific attention, than men. These gaps are believed to be  shaped by caregiving responsibilities that limit productivity~\cite{morgan2021unequal,stoet2018gender}, differences in research topic selection~\cite{erosheva2020nih,maddi2021gender}, outlet visibility~\cite{ross2020leaky}, and self-promotion~\cite{moss2010disruptions}, as well as explicit and implicit biases~\cite{rossiter1993matthew,moss2012science,knobloch2013matilda,teich2021citation}.

While cumulative advantage and gender disparities are both known to skew citations, they have been studied separately, leaving their  interplay largely underexplored. This gap matters: unequal distributions of attention shape careers, hiring, and funding, while systematically undervaluing women and other underrepresented groups. With web-mediated infrastructures like academic search engines and citation platforms increasingly determining whose work is seen, failing to address these dynamics risks amplifying inequities of the web of science.

We address this gap through two main contributions. First, we  quantify disparities in citation networks using large-scale bibliometric data from diverse scientific disciplines. 
Second, we introduce  a generative model of the growth of citation networks that explains how these disparities emerge and persist. This model, which incorporates key factors such as preferential attachment (cumulative advantage), homophily (biased citation preferences), unequal group size (underrepresentation), is analytically tractable yet captures real-world dynamics. 

By analyzing this model, we demonstrate that citation disparities  arise not only from underrepresentation but also from biased citation preferences and unequal integration of new authors into networks. Our model also replicates empirically observed disparities across scientific fields, revealing how both gender and institutional prestige influence citation patterns. The model also reproduces empirically observed variations in elitism across disciplines, where recognition becomes increasingly concentrated among a select group of authors from more prestigious institutions.

Together, our empirical analysis and modeling reveal why increasing representation alone is insufficient, and point instead to structural interventions, such as reducing homophily, amplifying contributions from underrecognized groups, and ensuring equitable integration of newcomers, as necessary for building a more fair, transparent, and responsible web of science.

\vspace{-0.3cm}
\section{Related Work}
\textbf{Cumulative advantage.}
Inequalities are embedded in the structure of science, with recognition and rewards concentrated among a small set of top performers~\cite{merton1968matthew,allison1982cumulative,nielsen2021global}. This concentration is often explained by the ``rich get richer'' mechanism of cumulative advantage~\cite{merton1968matthew}, whereby already distinguished researchers attract disproportionate attention compared to their less advantaged peers. While sometimes framed as beneficial to science by stratifying researchers according to merit~\cite{xie2014undemocracy}, this dynamic reinforces advantage by granting the most visible researchers privileged access to high-profile collaborators~\cite{nielsen2021global}, mentors~\cite{sekara2018chaperone}, funding, and other factors~\cite{bol2018matthew,kong2022influence}, which in turn further amplify their recognition and as a result, citations.

Recent evidence shows that factors beyond merit significantly influence citation patterns. For instance, papers published in a now-defunct journal were cited 20\% less often than comparable papers in active journals~\cite{rubin2021systematic}.

\vspace{0.1cm}
\noindent
\textbf{Gender disparities.} Across disciplines such as neuroscience~\cite{Dworkin2020}, political science~\cite{dion2018gendered}, astronomy~\cite{caplar2017quantitative}, and physics~\cite{teich2021citation}, women consistently receive fewer citations than men for similar work. These gaps cannot be explained solely by underrepresentation but are thought to reflect caregiving responsibilities~\cite{morgan2021unequal}, research topic choice~\cite{erosheva2020nih,maddi2021gender}, and differences in publication outlets~\cite{ross2020leaky} or self-promotion~\cite{moss2010disruptions}. Biases, both explicit and implicit, further devalue women’s contributions~\cite{rossiter1993matthew,moss2012science,knobloch2013matilda,teich2021citation}. However, the interaction between these various forms of bias and mechanisms of cumulative advantage has not been deeply explored.

\vspace{0.1cm}
\noindent
\textbf{Network structures.} Prior research shows that social networks can systematically disadvantage minorities by reducing their visibility~\cite{lee2019homophily}, relative ranking~\cite{karimi2018homophily,espin2022inequality}, economic opportunities~\cite{jackson2021inequality}, and number of connections~\cite{avin2015homophily,avin2017modeling}. However, most of this work focuses on undirected networks (collaborations, affiliations). Our focus on directed citation networks uncovers distinct mechanisms of disparity, with attention asymmetries shaping career trajectories and scientific influence.

\section{Power-Disparity in Citation Networks}
\label{sec:power_inequality}
As our first contribution, we define a measure of citations disparity in author-citation networks.
An author-citation network at time $\timeValue = 1,2, \ldots$ is a directed graph $\graph_\timeValue = \{\nodeSet_\timeValue, \edgeSet_\timeValue\}$ where $\nodeSet_\timeValue$ is the set of authors, and $\edgeSet_\timeValue$ is the set of directed edges between them. An edge from author $u$ to $v$ exists if the latter cites any of the papers written by the former. Hence, the out-degree $\outDegree(v)$ of an author $v \in \nodeSet_\timeValue$ is the number of authors citing $v$, and the in-degree $\inDegree(v)$ is the number of people $v$ has cited up to and including time $\timeValue$. We partition the set of authors into two groups $\calBlue_\timeValue, \calRed_\timeValue \subset \nodeSet_\timeValue$ based on gender (or alternately, institutional prestige),
with $\calRed_\timeValue$ representing the minority female authors and $\calBlue_\timeValue$ the majority male authors. 
The total in-degree (resp. out-degree) of red nodes is $\inDegree(\calRed_\timeValue) = \sum_{v \in{\calRed_\timeValue}}\inDegree(v)$  (resp. $\outDegree(\calRed_\timeValue) = \sum_{v \in{\calRed_\timeValue}}\outDegree(v)$), and a similar notation is used for blue nodes as well. Further, the total in-degree of all nodes at time $\timeValue$ (also equal to the total out-degree of all nodes at time $\timeValue$) is $\degree_\timeValue = \sum_{v \in V_\timeValue} \inDegree(v)$.

\begin{figure*}
    \vspace{-0.5cm}
        \centering
        \hspace{0.5cm}
        \begin{subfigure}[b]{0.44\textwidth}
        \includegraphics[width=\textwidth]{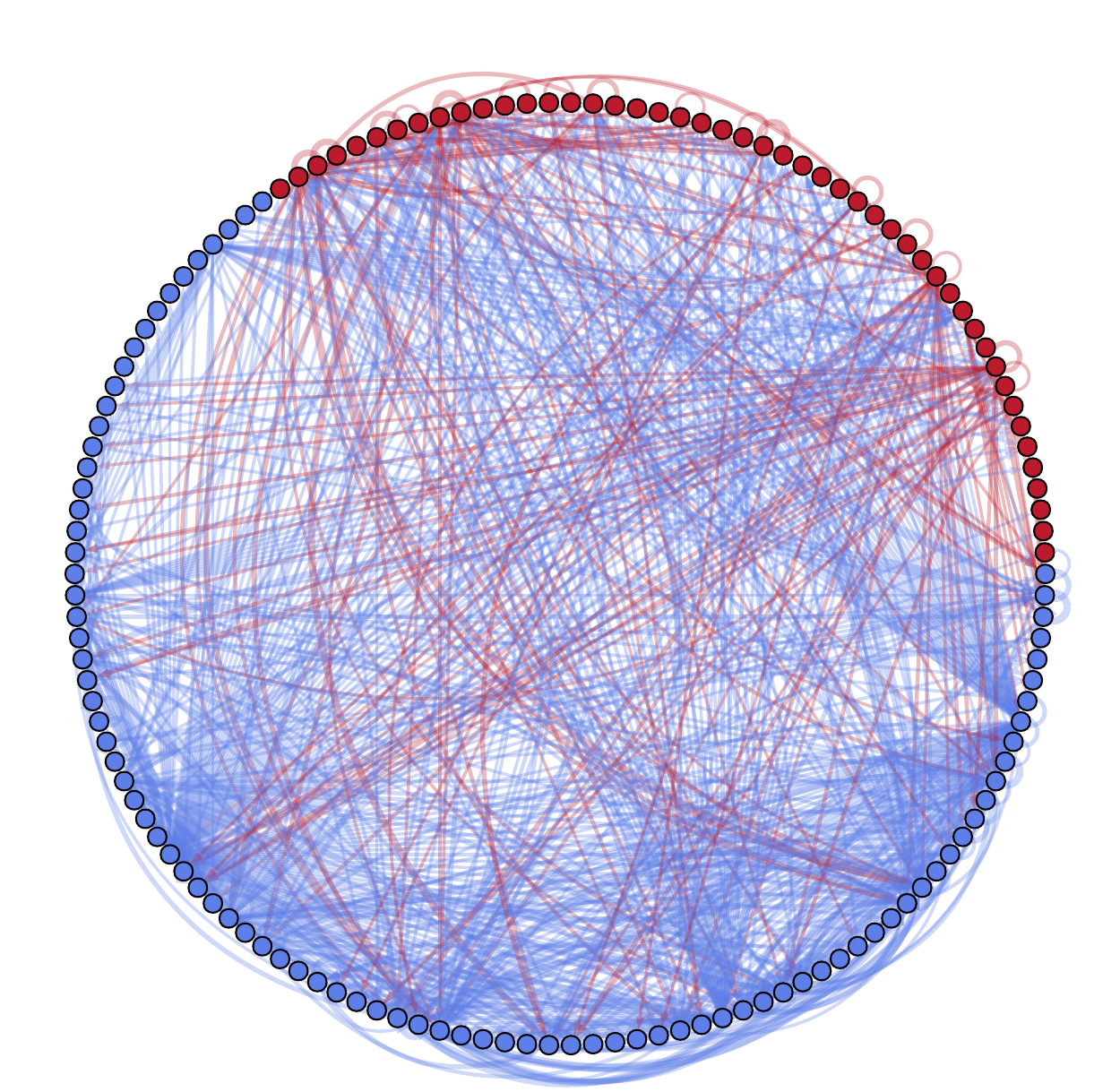}
            \caption[]%
            {A subgraph of gender-partitioned network}
        \end{subfigure}
        \hspace{1.1cm}
        \begin{subfigure}[b]{0.44\textwidth}
            \centering
        \includegraphics[width=\textwidth]{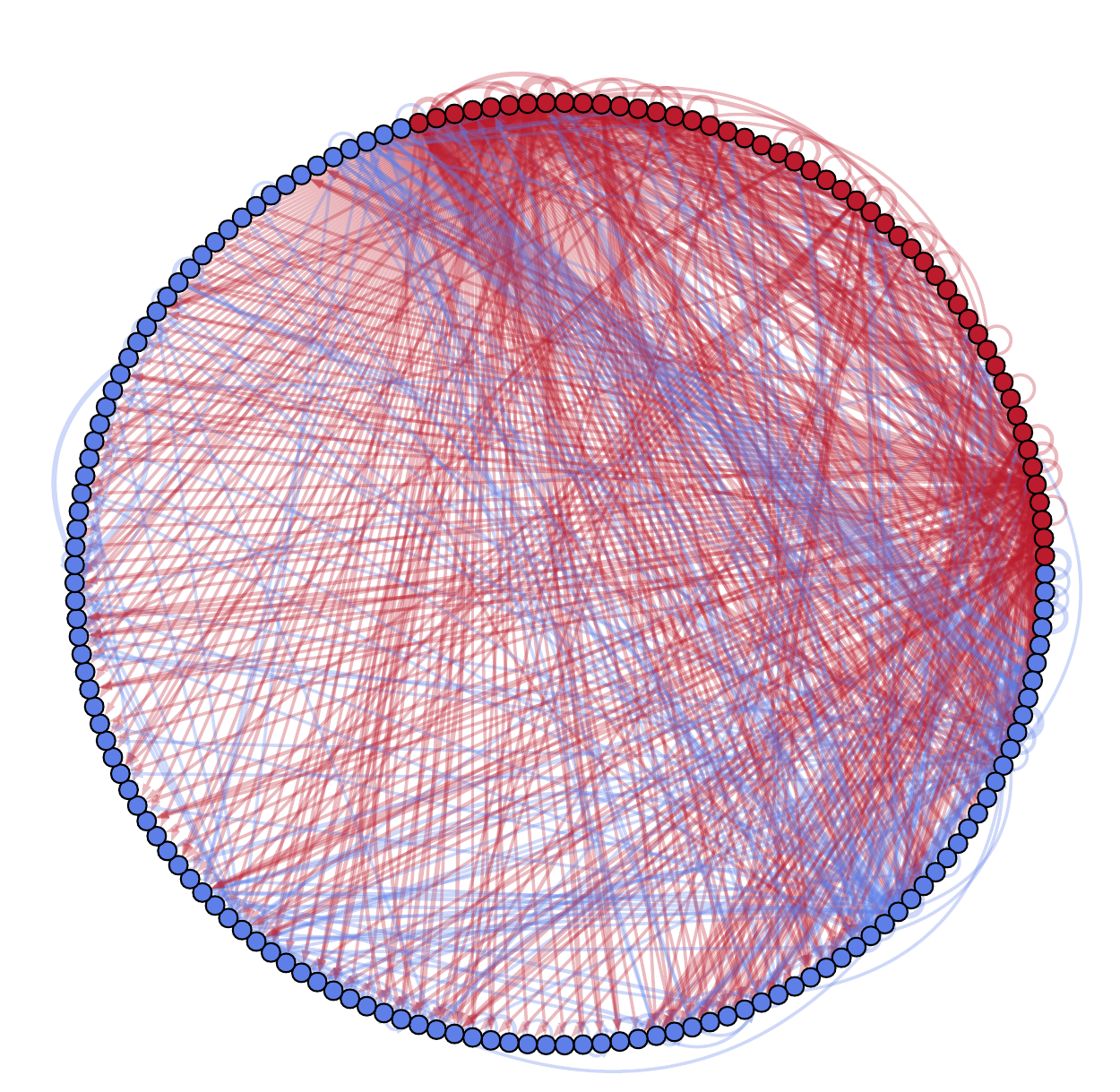}
            \caption[]%
            {A subgraph of the affiliation-partitioned network}
        \end{subfigure}           
        \hfill
        \begin{subfigure}[b]{0.48\textwidth}
        \vspace{0.2cm}
        \includegraphics[width=\textwidth]{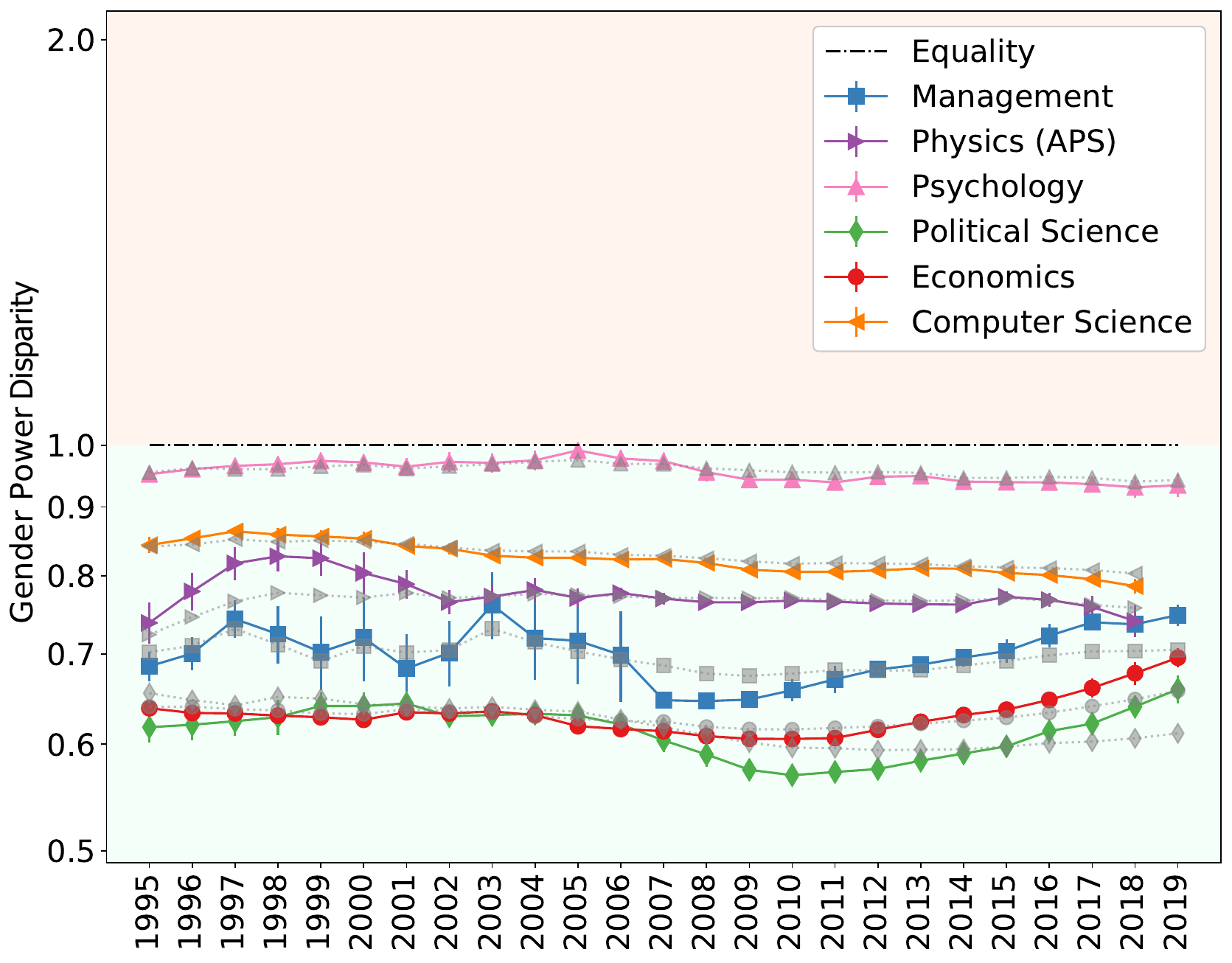}
            \caption[]%
            {{\small Gender Power-disparity of full network over time}}
        \end{subfigure}
        \hfill
        \begin{subfigure}[b]{0.48\textwidth}
        \vspace{0.2cm}
        \includegraphics[width=\textwidth]{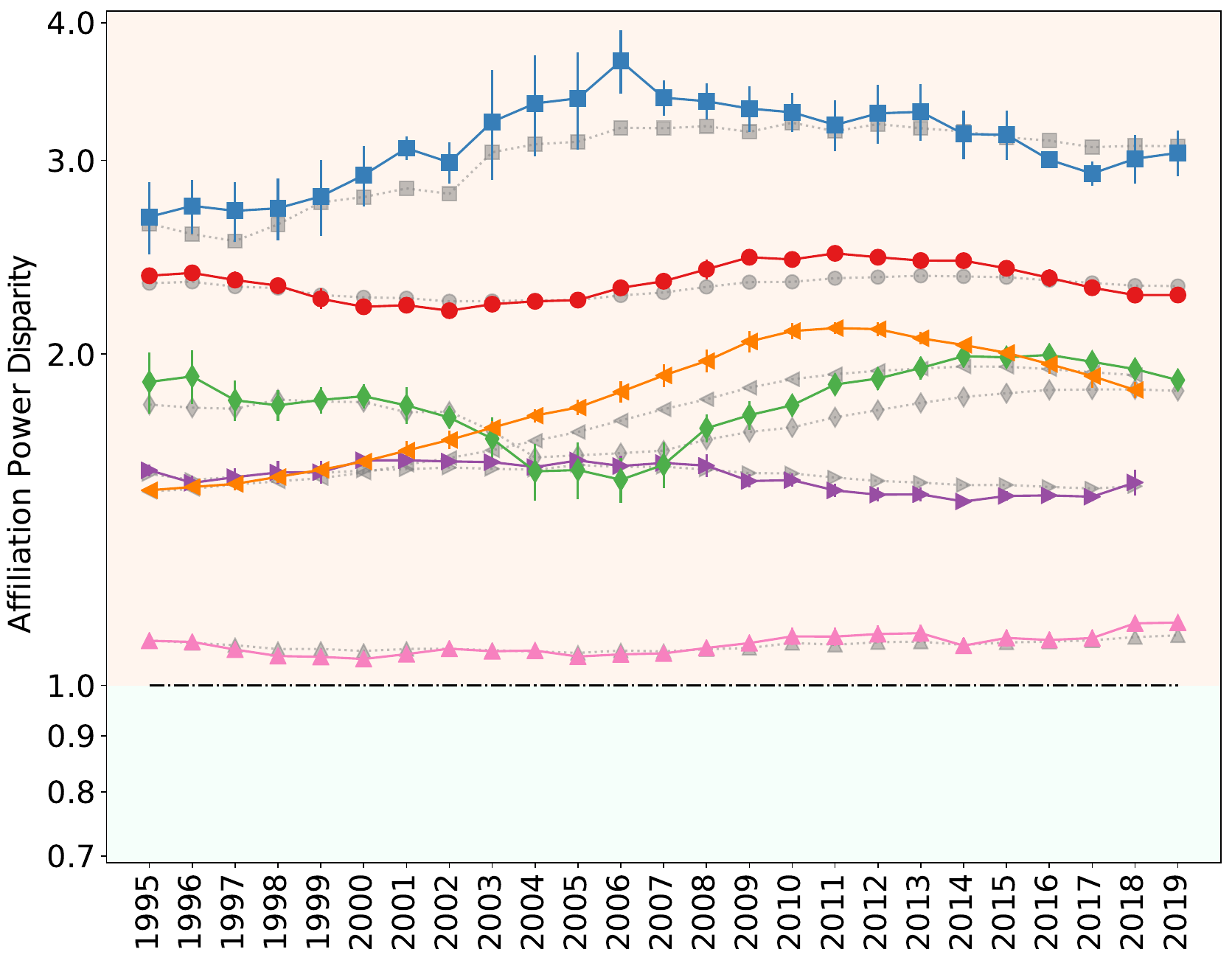}
            \caption[]%
            {{ Affiliation Power-disparity of full network over time}}
        \end{subfigure} 
        \hfill
        \vspace{-0.2cm}
        \caption{ 
       A subgraph of the Management citation network partitioned by \textbf{(a)} gender and \textbf{(b)} institutional prestige. The subgraph is constructed by sampling a linked pair of nodes and including all nodes that cite or are cited by them, along with their edges. In (a), nodes are partitioned by gender (red are women); in (b), they represent authors from top-100 vs. other institutions (Shanghai Rankings). Edges take the color of the cited author. The gender-partitioned network shows disproportionate citations to the majority male group, while the prestige-partitioned network shows disproportionate citations to the minority elite group. The lower panels plot power-disparity (Eq.~\eqref{eq:power_inequality}) over time for six large fields. Values are averaged over a four-year sliding window, with confidence intervals showing standard error. In (a), values consistently below 1 indicate women hold less power than men; in (b), values consistently above 1 indicate top-ranked institutions hold more power despite being a minority. Appendix~\ref{appn:empirical_details} provide additional details on our datasets.
        } 
              \label{fig:power}
\end{figure*}

Following \cite{avin2015homophily}, we use the average degree of a group as a proxy of its power in 
a graph. However, in directed graphs, we need to account for the in-degree $\inDegree(v)$ and the out-degree $\outDegree(v)$ for each author $v$.
We define the power of the red group~$\calRed_\timeValue$ 
at time $\timeValue$ as the ratio of the average out-degree to the average in-degree of red nodes at time $\timeValue$, and similarly for the blue group.  
In other words, the power of each group measures the average recognition the group receives relative to the average recognition the group gives to others. 
The power-disparity defined below quantifies the imbalance in recognition the two groups receive.

\begin{definition}[power-disparity]
\label{def:power_disparity}
    The power-disparity $\powerInequality_\timeValue$ in the citation network at time $\timeValue$ is the ratio of the power of the red group to the power of the blue group, namely,~
\begin{equation}
	\label{eq:power_inequality}
	\powerInequality_\timeValue 
	= \frac{\outDegree(\calRed_\timeValue) }{\inDegree(\calRed_\timeValue)} \times \frac{\inDegree(\calBlue_\timeValue) }{\outDegree(\calBlue_\timeValue)}.
\end{equation}
\end{definition}

When $\powerInequality_\timeValue <1$, the
red group has less power  relative to the blue group at time $\timeValue$ whereas $\powerInequality_\timeValue > 1$ indicates that the red group has relatively more power than the blue group. There is no power disparity when $\powerInequality_\timeValue  = 1$.


We define power-disparity (Eq.~\eqref{eq:power_inequality}) to capture inequalities between two non-overlapping groups (e.g., men vs. women, elite vs. non-elite) in directed citation networks. Unlike standard measures (e.g., variance, Gini), it accounts for both in-degree (citations given) and out-degree (citations received), is robust to self-citations\footnote{Note that each self-citation increases both the total out-degree and total in-degree of the same group by one unit. Thus, its effect is largely canceled out when the ratio of total out-degree to total in-degree is used as power a group.}, and yields no disparity under full reciprocity, ensuring it reflects true asymmetries rather than group size effects. It is simple to compute, requiring only degree counts and group affiliation, yet remains theoretically tractable and empirically effective, revealing real-world disparities while enabling rigorous analysis of their mechanisms.

Fig.~\ref{fig:power} illustrates power-disparity in large-scale bibliometric data (see Table~\ref{tab:data_info} in the appendix for summary statistics of our datasets). Panels (a)–(b) show citation subgraphs in Management science, partitioned by gender and affiliation prestige. Outgoing edges inherit the color of the author being cited. Despite similar group sizes, citation flows differ: in (a), the majority male group (blue) receives disproportionate share of citations, while in (b), the minority elite group (red) does. Panels (c)–(d) confirm these dynamics at scale across six fields. In gender-partitioned networks, women (the minority $\calRed_{\timeValue}$) consistently hold less power than men, with Political Science and Economics showing the largest disparities and Psychology coming closest to parity. It is also most gender balanced, while in Political Science and Economics, women are a minority with 34.20\% and 28.01\% of all authors.

Class imbalance alone does not explain citation disparities. Partitioning authors by institutional prestige shows that top-100 institutions, though a minority, hold far more power than non-elite institutions. Elite authors consistently receive greater recognition across fields (Fig.~\ref{def:power_disparity}(d)), with Economics and Management showing the largest gaps, while Psychology is closest to parity for both prestige and gender.

Strikingly, our symmetric definition of power-disparity reveals consistent imbalances: female authors and those from non-elite institutions hold less power across fields and time. Since group size alone cannot explain these patterns, key questions arise: what sociological factors drive these disparities, why do they vary by field, and what interventions might reduce them?

\section{A Dynamical Model of Network Growth}

	To explore the origins of the observed citations disparities, we present a model of a growing bi-populated network. It captures the key elements of real-world dynamics, while being simple enough for theoretical analysis.
	The proposed model has parameters defining class balance ($\redBirthProb$), growth dynamics of the network ($\probEventOne,\ \probEventTwo$), homophily ($\matrixEventOne,\ \matrixEventTwo , \matrixEventThree$) and preferential attachment ($\deltaIn,\ \deltaOut$). The model parameters $\matrixAnyEvent _i, i \in \{1,2,3\}$ are  matrices of the form
	\begin{equation}
		\matrixAnyEvent_{i} = \begin{bmatrix}
			\homophilyBlueAnyEvent^{(i)} & 1 - \homophilyRedAnyEvent^{(i)} \\
			1 - \homophilyBlueAnyEvent^{(i)} & \homophilyRedAnyEvent^{(i)} 
		\end{bmatrix}.
	\end{equation} 
	
	The network at time $\timeValue$ is denoted by $\graph_\timeValue = (\nodeSet_\timeValue, \edgeSet_\timeValue)$ where the set of nodes $\nodeSet_\timeValue$ can be partitioned into a set of blue nodes~$\calBlue_\timeValue$ and red nodes~$\calRed_\timeValue$. 
	The initial graph $\graph_0$ corresponds to $2\times 2$ adjacency matrix containing all 1's, though $\graph_0$ could be any arbitrary matrix and the asymptotic state of the model does not depend on it.
	
	\begin{figure}
		\centering
		\includegraphics[width=0.9\linewidth]{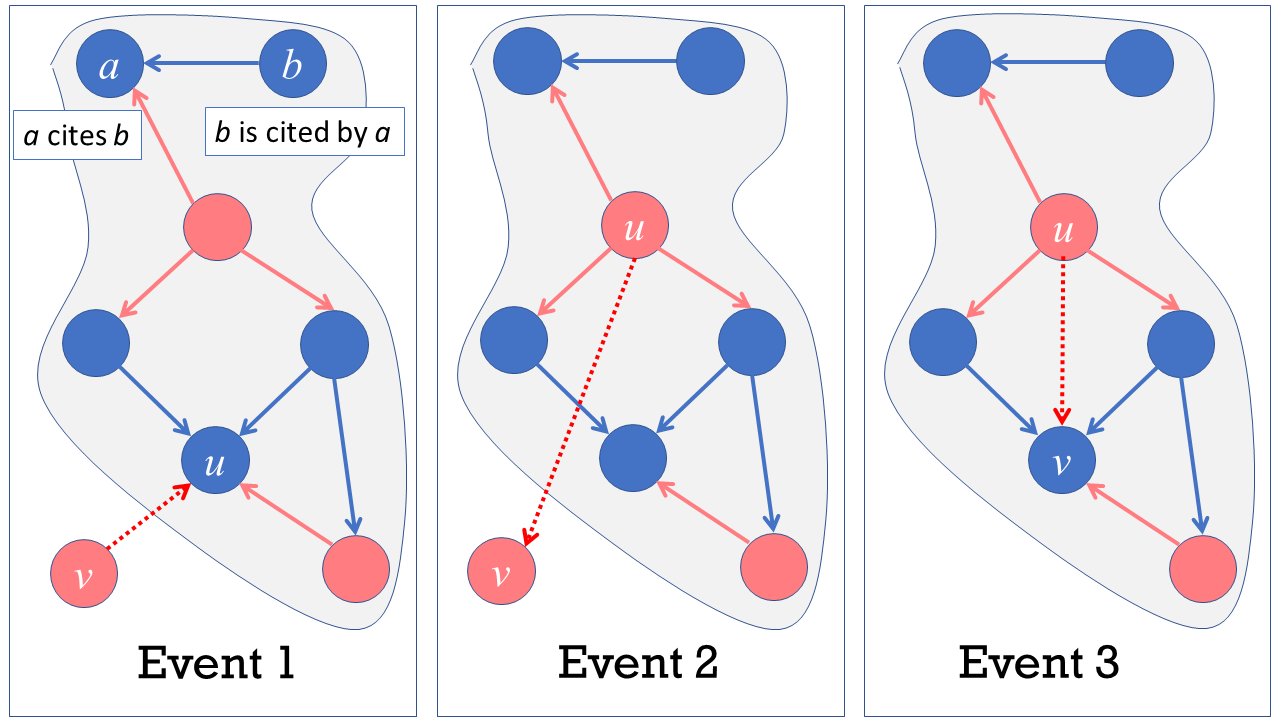}
        \vspace{-0.2cm}
		\caption{Citation edge creation events considered in the proposed DMPA model. The first two events correspond to the appearance of a new node that is either cited by (Event~1) or cites an existing node. In Event~3 (densification) an existing node cites another existing node. 
		}
		\label{fig:model_main}
	\end{figure}
	
	At each time step $\timeValue$, one of the following three events occur (as shown in Fig.~\ref{fig:model_main}):
	\begin{enumerate}[itemsep=0.1em, parsep = 0.2em]
		\item Event 1 (with probability $\probEventOne$): An existing node follows the new node
		\begin{enumerate}[label=\roman*., itemsep=0.05em, topsep=0pt, parsep = 0.2em]
			\item {\bf Minority - majority partition:} A new node $v_\timeValue$  appears and is assigned color red with probability $\redBirthProb$ and blue with probability $1-\redBirthProb$.
			
			\item {\bf Preferential attachment:} 
			An existing node $u_\timeValue \in \nodeSet_\timeValue$ is chosen by sampling with probability $\propto \inDegree(u_\timeValue) + \deltaIn$. 
			
			\item {\bf Homophily:} If both $u_\timeValue, v_\timeValue$ are red (resp. blue), an edge $(v_\timeValue, u_\timeValue)$ is added with probability $\homophilyRedEventOne$ (resp. $\homophilyBlueEventOne$). Otherwise, if $u_\timeValue$ is red (resp. blue) and $v_\timeValue$ is blue (resp. red), a link $(v_\timeValue, u_\timeValue)$ is added with probability $1 - \homophilyRedEventOne$ (resp. $1 - \homophilyBlueEventOne$). 
			
			\item Above steps ii, iii are repeated until an outgoing edge is added to $v_\timeValue$. 
		\end{enumerate}

		\item Event 2 (with probability $\probEventTwo$): The new node follows an existing node
		\begin{enumerate}[label=\roman*., itemsep=0.05em, topsep=0pt, parsep = 0.2em]
			\item {\bf Minority - majority partition:} as in case~1 above. 
			
			\item {\bf Preferential attachment:} An existing node $u_\timeValue \in \nodeSet_\timeValue$ is chosen by sampling with probability  $\propto \outDegree(u_\timeValue) + \deltaOut$. 
			
			\item {\bf Homophily:} If both $u_\timeValue, v_\timeValue$ are red (resp. blue), an edge $(u_\timeValue, v_\timeValue)$ is added with probability $\homophilyRedEventTwo$ (resp. $\homophilyBlueEventTwo$). Otherwise, if $u_\timeValue$ is red (resp. blue) and $v_\timeValue$ is blue (resp. red), a link $(u_\timeValue, v_\timeValue)$ is added with probability $1 - \homophilyBlueEventTwo$ (resp. $1 - \homophilyRedEventTwo$). 
			
			\item Above steps ii, iii are repeated until an incoming edge is added to $v_\timeValue$.
		\end{enumerate}

		\item Event 3 (with probability $1 - \probEventOne  - \probEventTwo$): Densification through new edges between existing nodes
		\begin{enumerate}[label=\roman*., itemsep=0.05em, topsep=0pt, parsep = 0.2em]
			\item {\bf Preferential attachment:} An existing node $u_\timeValue \in \nodeSet_\timeValue$ is chosen by sampling with probability  $\propto \outDegree(u_\timeValue) + \deltaOut$ and another node $v_\timeValue \in \nodeSet_\timeValue$ is chosen by sampling with probability  $\propto \inDegree(v_\timeValue) + \deltaIn$.

			\item {\bf Homophily:} as above in case (2) (but with $\homophilyRedEventThree$, $\homophilyBlueEventThree$). 
			
			\item Above steps i, ii are repeated until a new edge $(u_\timeValue, v_\timeValue)$ is added to the graph.
		\end{enumerate}
	\end{enumerate}

The probability $\redBirthProb$ that a new node is assigned to the red group
determines class balance asymptotically: when $\redBirthProb < 0.5$, red is the minority; otherwise, it is the majority.  For simplicity, we assume that homophily for the three edge creation events is the same and denoted by $\homophilyBlueAnyEvent, \homophilyRedAnyEvent$. We can independently set the minority and majority groups to be homophilic or heterophilic by adjusting  the parameters $\homophilyRedAnyEvent,\ \homophilyBlueAnyEvent$. If $\homophilyBlueAnyEvent > 0.5$, the blue group is homophilic,~i.e.,~blue nodes more likely to cite other blue nodes. On the other hand, if $\homophilyBlueAnyEvent < 0.5$, the blue group is heterophilic~i.e.,~blue nodes are more likely to cite red nodes. A similar interpretation holds for the red group (with the parameter~$\homophilyRedAnyEvent$).  We can control the growth dynamics of the graph by appropriately choosing $\probEventOne, \probEventTwo$, which determine the relative frequency with which new authors arrive and link to existing authors. For example, setting $1-\probEventOne-\probEventTwo > \probEventTwo > \probEventOne$ describes an author-citation network where most of the new citations form between existing authors, and less frequently when new authors cite (or are cited by) existing authors. Finally, $\deltaboth > 0$ (where we assume $\deltaIn = \deltaOut = \deltaboth$ for simplicity) determines
the degree of preferential attachment: larger $\deltaboth$ corresponds to 
lower preference for linking to high-degree nodes. {Further, the model also allows the formation of multiple edges among the nodes to indicate multiple citations among them.}

These parameters enable our proposed DMPA model to asymptotically replicate structural properties of many real-world social networks, including
scale free degree distribution~\cite{barabasi1999emergence}, assortative mixing~\cite{mcpherson2001birds,newman2002assortative}, and, as we show here, power-disparities.
The model generalizes previous models, specifically, a preferential attachment model for directed networks populated with nodes of a single type~\cite{bollobas2003directed} and a dynamic model for bi-populated undirected networks~\cite{avin2020mixed,avin2015homophily}.

{The proposed model generates a network with authors~(instead of papers) as nodes and citations as edges among authors. When a paper with multiple authors cites another paper with multiple authors, this can be 
viewed
as implementing the model at a faster time scale, with no changes to the asymptotic state of the model.}

\subsection{Theoretical Analysis of the Model} 
\label{subsec:theoretical_analysis}

\begin{figure*}[!h]
\vspace{-0.2cm}
	\includegraphics[width=\textwidth,  trim={0.15cm 0.3cm 0.2cm 0.1cm},clip]{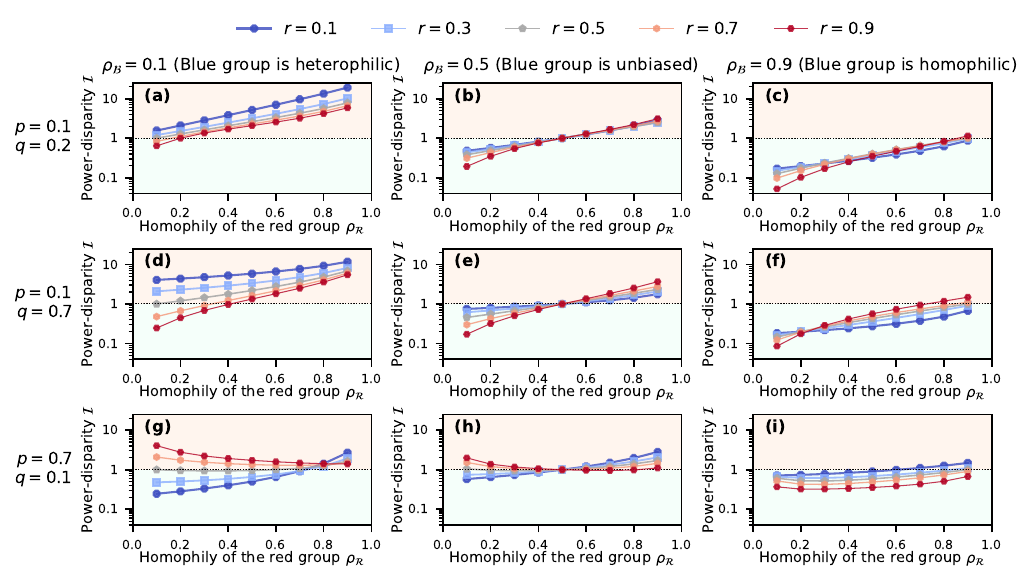}
    \vspace{-0.4cm}
	\caption{
		The figures display the power-disparity values for various parameter regimes of the DMPA model.
		The three rows correspond to different values of the parameters $\probEventOne, \probEventTwo$ that capture the growth dynamics of the DMPA model. The three columns correspond to three different values of the homophily parameter of the blue group:~$\homophilyBlueAnyEvent = 0.1$~(heterophilic), $\homophilyBlueAnyEvent = 0.5$~(unbiased), $\homophilyBlueAnyEvent = 0.9$~(homophilic). In each subplot, lines in different colors correspond to various values of the parameter $\redBirthProb$ that determines the asymptotic class balance. In addition, $\deltaboth = 10$ for each case and another figure for $\deltaboth = 100$~(Fig.~\ref{fig:PowerInequalityTheoretical_delta100}) is given in the Appendix~\ref{appn:additional_results}. The power disparity values are computed using the recursive method based on Theorem~\ref{th:convergence_DMPA} (discussed in Sec.~\ref{subsec:theoretical_analysis}).
	} 
	\label{fig:PowerInequalityTheoretical}
\end{figure*}

We analyze the proposed DMPA model to study the asymptotic values of power-disparity (Definition~\ref{def:power_disparity}). First, we define some additional notation. Let 
\begin{equation}
	\label{eq:thetaTime}
	\thetaIn_\timeValue = \frac{\inDegree(\calRed_\timeValue)}{\degree_\timeValue}, \quad \thetaOut_\timeValue = \frac{\outDegree(\calRed_\timeValue)}{\degree_\timeValue}
\end{equation}
represent the fraction of the total in-degree and out-degree of the red group at time $\timeValue$ {as a fraction of the total in-degree of all
nodes at time $\timeValue$.}

\begin{theorem}[Almost sure convergence of the state in DMPA model]
	\label{th:convergence_DMPA}
	Consider the DMPA model with the parameters $\redBirthProb, \probEventOne , \probEventTwo$, $\deltaIn = \deltaOut = \deltaboth$ and $\homophilyBlueAnyEvent^{(i)} = \homophilyBlueAnyEvent, \homophilyRedAnyEvent^{(i)} = \homophilyRedAnyEvent$ for $i  = 1,2,3$. Then, there exists $\deltaboth^{*} > 0$ such that, for all $\deltaboth > \deltaboth^{*}$, the state of the system $\thetaVec_\timeValue = [\thetaIn_\timeValue, \thetaOut_\timeValue]^T$
	converges to a unique value ${\thetaVec_{*} = [\thetaIn_{*}, \thetaOut_{*}]^T \in [0,1]^2}$  with probability $1$ as time $\timeValue \rightarrow \infty$.
    \vspace{-0.1cm}
\end{theorem}
Theorem~\ref{th:convergence_DMPA} states that the normalized sum of in-degrees~$\thetaIn_\timeValue$ and the normalized sum of out-degrees~$\thetaOut_\timeValue$  of red nodes converge to unique values (for sufficiently large $\deltaboth$) with probability~$1$. We can then utilize these unique asymptotic values to analyze the power-disparity. For simplicity, we have assumed that the homophily parameters do not differ across the three edge formation events and $\deltaIn = \deltaOut$, although this assumption can be easily relaxed.  

The full proof of Theorem~\ref{th:convergence_DMPA} is given in Appendix~Sec.~\ref{SI:proof_of_convergence}. The key idea of the proof is to first express the evolution of $\thetaVec_\timeValue = [\thetaIn_\timeValue, \thetaOut_\timeValue]^T$ as a non-linear stochastic dynamical system. Then, by using stochastic averaging theory, we show that the non-linear stochastic dynamical system can be approximated by a deterministic ordinary differential equation~(ODE) of the form $\Dot{\thetaVec} = \nonLinFunction(\thetaVec) - \thetaVec$ where $\nonLinFunction$ is a contraction map~(for sufficiently large $\deltaboth$). Since $\nonLinFunction$ is a contraction map, it has a unique fixed point and the sequence~$\thetaVec_\timeValue, \timeValue = 1, 2, \ldots$ converges to that fixed point (by the Banach fixed point theorem). Since we know the contraction map $\nonLinFunction$ in closed form~(see Appendix~Sec.~\ref{SI:proof_of_convergence}), the unique fixed point~$\thetaVec_{*} = [\thetaIn_{*}, \thetaOut_{*}]^T$ can be computed easily via the recursion $\thetaVec_{k+1} = \nonLinFunction(\thetaVec_{k})$ with arbitrary initial condition $\thetaVec_{0} \in [0,1]^2$. Thus, Theorem~\ref{th:convergence_DMPA} implies that the power-disparity $\powerInequality_\timeValue$ in Eq.~\ref{eq:power_inequality} converges to 
$\powerInequality = {\thetaOut_{*}(1-\thetaIn_{*})}/{\thetaIn_{*}(1-\thetaOut_{*})}$. In addition, the asymptotic value of the power-disparity $\powerInequality$ for a specific parameter configuration can be found iteratively.
%
In contrast to \cite{avin2015homophily, avin2020mixed, bollobas2003directed}, which established convergence of $\thetaVec_\timeValue$ in expectation (for different models), we use tools from stochastic averaging theory to establish the almost sure convergence~(i.e.,~convergence with probability~$1$). 

\subsection{Insights from the  Theoretical Analysis 
}
\label{subsec:insights_theoretical_analysis}
\begin{figure*}[h!] 
        \centering
        \begin{subfigure}{0.48\textwidth}
            \centering
            \includegraphics[width=\textwidth]{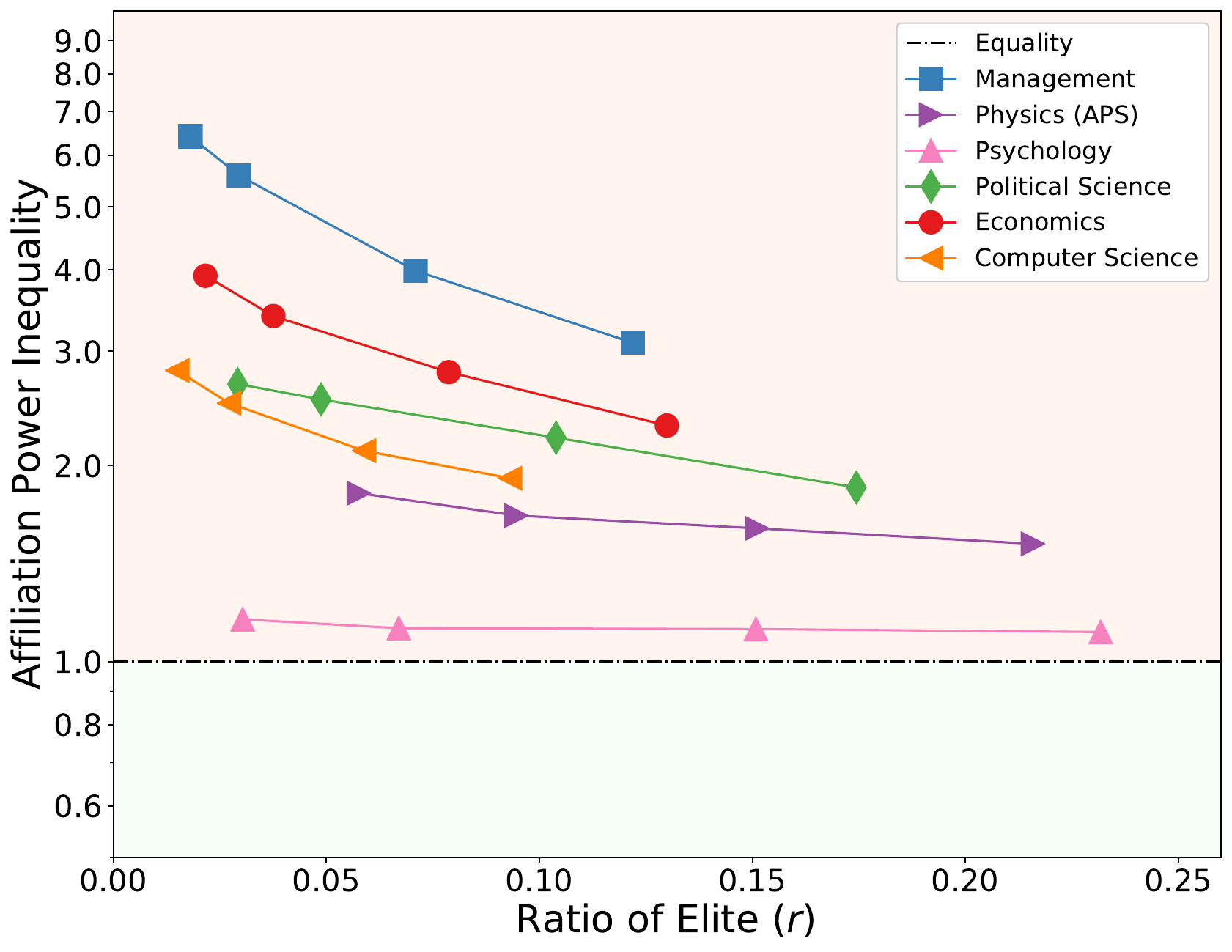}
            \caption{Prestige power-disparity}
        \end{subfigure}
        \hfill
        \begin{subfigure}{0.48\textwidth}
        \includegraphics[width=\textwidth]{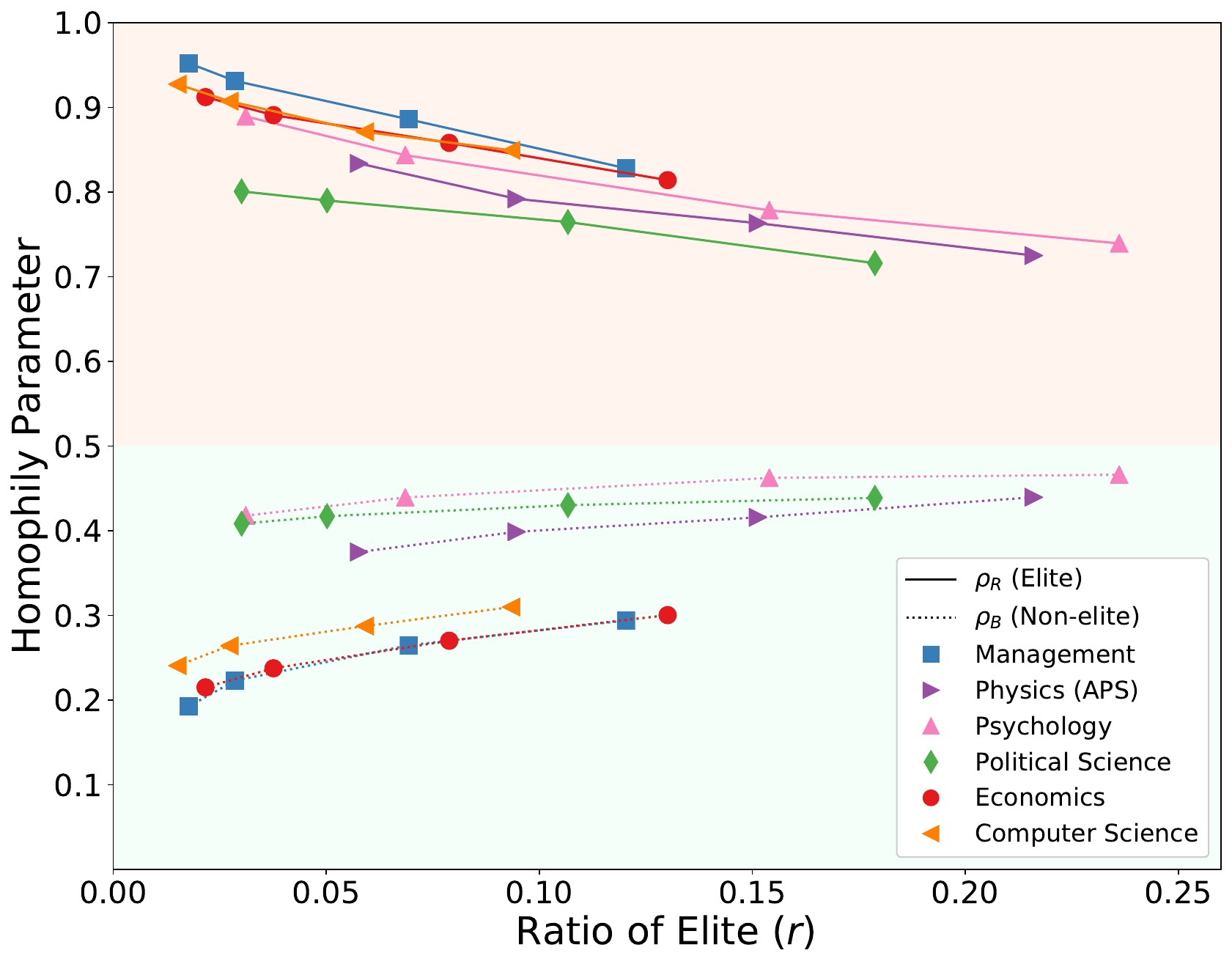}
            \caption{Homophily parameters}
        \end{subfigure}
        \vspace{-0.3cm}
    \caption{The figures illustrate how (a)~power inequality and, (b)~homophily parameters of the two groups, vary with the minority group size in affiliation networks by defining ``top-ranked universities" to be the top 10, 20, 50 and 100 universities in \emph{Shanghai University Rankings (SUR).}
    We use $\deltaboth$ from Table \ref{tab:params} for each field of study. The figure shows that smaller the minority group is, the more powerful it is, as a consequence of minority and majority groups being increasingly homophilic and heterophilic, respectively. This observation agrees with the predictions of the DMPA model shown in Fig.~\ref{fig:PowerInequalityTheoretical}(a)
    \label{fig:elite_range}}
\end{figure*}

Fig.~\ref{fig:PowerInequalityTheoretical} shows asymptotic power-disparity values from the iterative method under different parameter settings, illustrating the range of behaviors generated by the DMPA model. 
Real-world networks typically densify~\cite{leskovec2007graph}, with new nodes citing existing ones but rarely the reverse.  
The top row in Fig.~\ref{fig:PowerInequalityTheoretical} illustrates this case, with  $\probEventOne = 0.1,\ \probEventTwo = 0.2$. When the blue group is heterophilic (Fig.~\ref{fig:PowerInequalityTheoretical}(a)), the red group gains power if it is sufficiently more homophilic. Here, power-disparity grows exponentially with red homophily but decreases with red group size, meaning smaller, highly homophilic red groups can dominate. Thus, when one group is strongly homophilic and the other strongly heterophilic, the heterophilic group faces persistent disadvantage regardless of size.
Therefore, even without altering preferential attachment, a group can increase its power by reducing its size or adjusting citation preferences. Intuitively, like an exclusive club whose members prefer one another and outsiders seek association, shrinking the group amplifies its influence. Fig.~\ref{fig:elite_range}(a) confirms this dynamic with the real-world citation networks: narrowing top-ranked affiliations to a smaller set of institutions increases both power-disparity and elite homophily.

Interestingly, our analysis suggests that an effective way to reduce the disparity in power of highly exclusive (i.e.,~$\redBirthProb \ll 1$), homophilic groups (i.e.,~$\homophilyRedAnyEvent\gg 0.5, \homophilyBlueAnyEvent\ll 0.5$) is to reduce the strength of preferential attachment by increasing $\deltaboth$. This can be seen by comparing Fig.~\ref{fig:PowerInequalityTheoretical}(a) with Fig.~\ref{fig:PowerInequalityTheoretical_delta100}(a) in Appendix~\ref{appn:additional_results}, calculated for $\deltaboth = 100$; note that reducing the red group size does not amplify its power as much as in the case of $\deltaboth = 10$ in Fig.~\ref{fig:PowerInequalityTheoretical}(a).
On the other hand, when the blue group is unbiased (Fig.~\ref{fig:PowerInequalityTheoretical}(b)),
the red group is more powerful 
when $\homophilyRedAnyEvent>\homophilyBlueAnyEvent$, 
and its power increases with group size.  There is less variation in power-disparity (compared to Fig.~\ref{fig:PowerInequalityTheoretical}(a)), suggesting that power disparity is largely driven by the difference of the homophily parameters of the two groups. {Further,  irrespective of the group size $\redBirthProb$, there is no disparity in power (i.e.,~$\powerInequality = 1$) when $\homophilyBlueAnyEvent = \homophilyRedAnyEvent = 0.5$. Therefore, the most effective approach for ameliorating power-disparity is to alter individual preferences by changing the homophily parameters~$\homophilyBlueAnyEvent, \homophilyRedAnyEvent$ to be neutral towards the groups.} Finally, when the blue group is highly homophilic (Fig.~\ref{fig:PowerInequalityTheoretical}(c))
it is more powerful than the red group~(i.e.,~$\powerInequality \leq 1$), 
and power-disparity decreases with~$\homophilyRedAnyEvent$.

The middle row of Fig.~\ref{fig:PowerInequalityTheoretical} ($\probEventOne < 1-\probEventOne-\probEventTwo < \probEventTwo$), represents a fast-growth phase, as in emerging fields where many new authors cite existing ones. Trends from the earlier setting persist, but group size ($\redBirthProb$) plays a stronger role due to the higher influx of new nodes. 

Finally, the bottom row ($\probEventOne > 1-\probEventOne-\probEventTwo > \probEventTwo$) models networks that grow mainly through existing authors citing new ones.

Extreme power-disparity ($\homophilyBlueAnyEvent = 0.1, \homophilyRedAnyEvent = 0.9$ or vice versa) lessens from top to bottom rows, suggesting that frequent entry of new authors can mitigate disparities under strong homophily. 

Our results so far examined power-disparity under strong preferential attachment ($\deltaboth = 10$;  Fig.~\ref{fig:PowerInequalityTheoretical}). 
To test its effect, we re-evaluated with weaker attachment ($\deltaboth = 100$; Fig.~\ref{fig:PowerInequalityTheoretical_delta100} in Appendix~\ref{appn:additional_results}). The comparison shows that preferential attachment amplifies disparities, especially under strong class imbalance $\redBirthProb \ll 0.5$ or $\redBirthProb \gg 0.5$, suggesting that reducing preferential attachment can mitigate inequality in such cases (Fig.~\ref{fig:PowerInequalityTheoretical}(a) vs. Appendix~Fig.~\ref{fig:PowerInequalityTheoretical_delta100}(a)).

\definecolor{Grey}{gray}{0.9}
\begin{table}[h!]
\small
	\centering
	\begin{tabular}{|c|c|c|c|c|c|c|}
		\hline
		\rowcolor{Grey}
		\multicolumn{7}{|c|}{\textit{Gender-partitioned Network Parameters}} \\
		\hline 
		\rowcolor{Grey}
		& \textit{Mgmt.
        } & \textit{Phys.} & \textit{Psych.} & \textit{Poli. Sci.} & \textit{Econ.} & \textit{Com. Sci.} \\
		\hline 
		$T$ & $1.19M$ & $17.23M$ & $12.79M$ & $7.24M$ & $94.40M$ & $435.66M$\\ 
		\hline 
		$\deltaboth$ & $1000$ & $1000$ & $4$ & $1000$ & $100$ & $20$ \\ 
		\hline 
		$r$ & $0.35$ & $0.16$ & $0.50$  & $0.34$ & $0.28$ & $0.26$\\ 
		\hline 
		$p$ & $0.025$ & $0.001$ & $0.030$  & $0.067$ & $0.004$ & $0.005$\\ 
		\hline 
		$q$ & $0.058$ & $0.008$ & $0.032$  & $0.064$& $0.009$ & $0.012$ \\ 
		\hline 
		$\homophilyRedAnyEvent$ & $0.46$ & $0.48$ & $0.54$ & $0.47$ & $0.48$ & $0.55$ \\
		\hline 
		$\homophilyBlueAnyEvent$ & $0.61$ & $0.62$ & $0.57$ & $0.67$ & $0.62$ & $0.57$ \\
		\hline
		$\powerInequality_{\mathrm{emp.}}$ & $0.70$ & $0.76$ & $0.94$ & $0.61$ & $0.66$ & $0.80$ \\
		\specialrule{.12em}{.1em}{.1em}
		$\powerInequality_{\mathrm{thr.}}$  & $0.74$ & $0.72$ & $0.82$ & $0.69$ & $0.65$ & $0.87$ \\ 
		\hline
	\end{tabular}
	\vspace{0.1in}    
	\caption{Estimated parameters of the DMPA model for the gender-partitioned networks. These values are also illustrated visually in Fig.~\ref{fig:estimated_parameters} (alongside corresponding values for affiliation-partitioned networks for comparison). Estimates are based on Microsoft Academic Graph data, except for Physics (APS), using the datasets and procedure described in the Appendix~\ref{appn:empirical_details}. 
    {Repeating the procedure multiple times yields parameter estimates with standard deviation less than $10^{-4}$}.}
	\label{tab:params}
\end{table}
\begin{figure}
    \centering\includegraphics[width=1.0\columnwidth,  trim={0.25cm 0.25cm 0.25cm 0.25cm},clip]{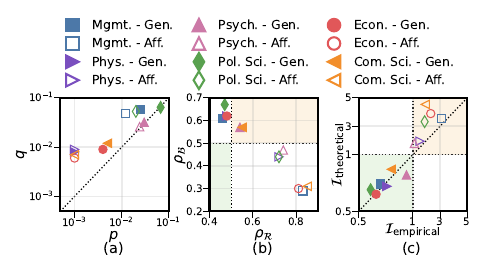}
    \vspace{-0.5cm}
    \caption{Empirically estimated DMPA parameters for gender-partitioned (filled markers) and affiliation-partitioned (open markers) networks; exact values are listed in Table~\ref{tab:params}~(gender-partitioned networks) and Appendix~Table~\ref{tab:params2}~(affiliation-partitioned networks). Subplot (a) shows that, across most fields, new nodes are more likely to cite existing authors than the reverse ($\probEventTwo > \probEventOne$). Subplot~(b) indicates that in gender-partitioned networks, minority female authors are typically heterophilic ($\homophilyRedAnyEvent < 0.5$) while the majority male authors are homophilic ($\homophilyBlueAnyEvent > 0.5$), with Psychology and Computer Science as exceptions where disparities are smaller. In affiliation-partitioned networks the pattern is reversed: elite authors (minority) are homophilic, while non-elite authors are heterophilic. Subplot (c) shows close agreement between empirical and model-based estimates of power-disparity, confirming that the DMPA model captures how disparities emerge in real-world networks and can inform strategies for mitigation.}
    \label{fig:estimated_parameters}
    \vspace{-0.5cm}
\end{figure}

\section{Fitting the Model to Real Networks}

To connect the DMPA model to real-world citation networks, we estimated model parameters from bibliographic data (see Appendix~\ref{appn:empirical_details} for details on our datasets and methods). Using these estimates, theoretical power-disparity ($\powerInequality_{\mathrm{theoretical}}$) was computed with the recursive method, while empirical power-disparity (denoted by $\powerInequality_{\mathrm{empirical}}$) was obtained directly from gender- and prestige-partitioned data.
Fig.~\ref{fig:estimated_parameters} shows the estimated parameters for all fields as well as a comparison of $\powerInequality_{\mathrm{theoretical}}$ and $\powerInequality_{\mathrm{empirical}}$; detailed results appear in Table~\ref{tab:params} (gender) and Appendix~Table~\ref{tab:params2} (prestige).

In gender-partitioned networks, theoretical and empirical values of power-disparity align closely, indicating the model effectively captures citation dynamics. Both confirm that women, as the minority group, face consistent recognition gaps. Across fields, network growth is dominated by densification ($\probEventOne < \probEventTwo \ll 1-\probEventOne-\probEventTwo$),  with new authors joining infrequently. Male authors are homophilic in all fields, while female authors are generally heterophilic except in Psychology and Computer Science.
Hence, gender-partitioned citation networks generally fall into the scenario captured in Fig.~\ref{fig:PowerInequalityTheoretical}(c). 
In fact, the male (blue) group's homophily and the female (red) group's heterophily play a crucial role in the emergence of power disparity, as demonstrated earlier using our proposed model. This insight is also important for devising strategies to reduce power disparity as we discuss subsequently. 

In affiliation-partitioned networks (Appendix~Table~\ref{tab:params2}, Fig.~\ref{fig:estimated_parameters}), the minority elite group holds more power, likely reflecting endogenous factors that link top institutions with higher citation rates. The close match between theoretical and empirical values confirms the model’s generalizability, while also showing that minority status alone does not determine power.
Although theoretically calculated values somewhat overestimate power-disparity, they have the same ranking as the empirical values in all fields except Computer Science and Management. Compared to gender-partitioned networks (filled markers in Fig.~\ref{fig:estimated_parameters}), the minority group is strongly homophilic~($\homophilyRedAnyEvent > 0.7$ in all fields), while the majority is heterophilic~($\homophilyBlueAnyEvent < 0.5$ in all cases).  Further, note that $\probEventOne < \probEventTwo \ll 1-\probEventOne-\probEventTwo$; hence, the affiliation-partitioned networks correspond to the scenario shown in Fig.~\ref{fig:PowerInequalityTheoretical}(a). 

The preferential attachment parameter $\deltaboth$ is 1000 for most networks, except in gender-partitioned Psychology, Economics, and Computer Science, and affiliation-partitioned Political Science. In these cases, smaller homophily differences suggest that preferential attachment, rather than homophily, drives power-disparity.

Based on our analysis, {Psychology} is the most egalitarian research field, with power-disparity values close to one, and some of the smallest gaps between homophily parameters for the minority and the majority groups. In addition, half of the authors in the gender-partitioned network belong to the red group (as well as quarter of the authors in the affiliation-partitioned network), a larger fraction compared to other fields.
In contrast, {Political Science} has the highest empirical gender disparity. The poor agreement between theoretical and empirical values of power-disparity for Computer Science could be due to the strong temporal variability of the citation networks in Computer Science seen in Fig.~\ref{fig:power}.

Correlations between gender and institutional prestige are negligible (Pearson correlation ranges from  $\gamma=-0.01$ to $r=0.04$, $p<0.001$), confirming that gender disparities are not confounded by the prestige of affiliation.

\subsection{Elitism: Smaller Elites Have More Power}
To demonstrate the generalizability of our approach, we estimated the prestige power-disparity 
by changing the definition of ``elite'' from the top-100 institutions to the top-50, top-20, and top-10 institutions {listed in \textit{Shanghai University Rankings}. Fig.~\ref{fig:elite_range} shows the results. As the number of institutions considered as ``elite'' shrinks, so does the minority fraction of authors belonging to those institutions ($\redBirthProb$). 
However, as the size of the elite shrinks, its relative power grows. This effect, which we call \textit{elitism}, is more pronounced in some fields than others. Psychology and Physics are more egalitarian than Management, Economics, and Computer Science,  where the power of authors from top-10 institutions is almost double that of authors from the top-100 institutions.
The estimated homophily parameters (Fig.~\ref{fig:elite_range}(b)) suggest that the smaller minorities are more homophilic and prefer to cite themselves more, whereas the majority group is more strongly heterophilic. These observations agree with the theoretical insights obtained from Fig.~\ref{fig:PowerInequalityTheoretical}; since the minority (elite group) is highly homophilic and the majority is heterophilic, reducing the size of the minority further increases its power. Further, Fig.~\ref{fig:elite_range}(a) also shows that ranking of the academic fields based on the power-disparities remain the same for different sizes of the minority group ($\redBirthProb$), with management corresponding to the largest prestige power-disparity and psychology the smallest.

\subsection{Mitigating Power-Disparity in Author-Citation Networks} 

The DMPA model reveals that simply increasing the size of the minority group does not reduce citation disparities, since power-inequality persists even when groups are equal in size ($\redBirthProb = 0.5$). A more effective strategy is reducing homophily, as shown in Fig.~\ref{fig:PowerInequalityTheoretical}: when groups cite beyond their own members, disparities diminish even under imbalance. This might explain why Psychology and Computer Science, despite their very different group balances, have the least gender power-disparity  among all fields~(i.e.,~power-disparity values closest to one as seen from Fig.~\ref{fig:estimated_parameters}(c)): 
{the majority~(male group) is least homophilic and $(\homophilyBlueAnyEvent,\homophilyRedAnyEvent)$ is closest to $(0.5,0.5)$ in those two fields~(see Fig.~\ref{fig:estimated_parameters}(b)).}
 Political Science, Economics, Management, and Physics exhibit greater gender disparity, with the minority group (women, red) being heterophilic and the majority group (men, blue) being homophilic in all four cases.  Merely adding more women under such conditions without first addressing homophily may even worsen disparities (as discussed in Sec.~\ref{subsec:insights_theoretical_analysis} using Fig.~\ref{fig:PowerInequalityTheoretical}). This observation underscores the need for interventions that reshape citation preferences and not just the group composition.

Reducing homophily can be encouraged by creating more cross-group visibility and interactions—through mentoring programs, workshops, symposia, and collaborations—that amplify the work of underrecognized groups and promote more equitable citation practices.

Additionally, recall that the worst-case power-disparity~(which occurs when difference between $\homophilyBlueAnyEvent$ and $\homophilyRedAnyEvent$ is close to 1) is largest in the top row and smallest in the bottom row of  Fig.~\ref{fig:PowerInequalityTheoretical}. Therefore, encouraging new authors to join a field of study (i.e.,~making $\probEventOne, \probEventTwo \gg 1-\probEventOne - \probEventTwo$) and especially, providing them the opportunity to have an audience among established authors~(i.e.,~$\probEventOne > \probEventTwo \gg 1-\probEventOne - \probEventTwo$) also reduces the  power-disparity. Practically, this can be achieved
through mentoring programs, events at academic conferences dedicated to highlighting
the work of junior researchers, like Doctoral Symposia, etc.

Finally, when neither homophily, growth dynamics, nor group size can be controlled, gender power-disparity may be reduced by weakening preferential attachment  (increasing $\deltaboth$). This can be achieved by providing early-career researchers with resources and opportunities not solely tied to past performance—for example, allocating modest support independent of citations or broadening access to speaking and panel invitations at academic events.

In comparison, in affiliation-partitioned networks (Appendix~Table~\ref{tab:params2}, Fig.~\ref{fig:estimated_parameters}), power-disparity stems largely from homophily: the majority is heterophilic while the minority is strongly homophilic. As Fig.~\ref{fig:PowerInequalityTheoretical}(a) and Fig.~\ref{fig:elite_range} show, smaller minorities amplify this imbalance. Thus, disparities are not due to cumulative advantage alone, but due to its interaction with homophily. These interactions create a feedback loop that amplifies even small differences in citation preferences. Mitigation strategies include reducing the salience of affiliation in publications, encouraging entry of new authors, and weakening preferential attachment. Because model parameters differ by field, no universal strategy exists; instead, estimating field-specific parameters allows the DMPA model to serve as a testbed for tailored interventions.

\section{Discussion and Conclusion}

Disparities in scientific impact via citations persist across many fields and can systematically limit career opportunities, particularly for women in disciplines where they remain underrepresented~\cite{way2016gender,porter2019women,chen2016female}. Notably, gender differences in citation patterns are visible even among highly distinguished scientists (e.g., NAS members), suggesting that disparities are not confined to early career stages~\cite{lerman2022gendered}.

To characterize these inequities in the inherently directed setting of citations, we quantified group-level imbalances in author--citation networks using a novel power-disparity measure that accounts for both citations received and citations given. Unlike standard inequality measures, it is robust to self-citations and yields no disparity under full reciprocity, ensuring it captures directional asymmetries in recognition.  Empirically, we found persistent gender gaps in power across fields and, separately, substantial prestige-based disparities in which a minority of authors from elite institutions can hold disproportionate power. 

To explain how such disparities emerge, we introduced a theoretically tractable model of citation-network growth, which integrates (i) unequal group representation, (ii) biased citation preferences (homophily/heterophily), (iii) preferential attachment (cumulative advantage), and (iv) field-specific growth dynamics. We prove that power-disparity converges to a unique value almost surely  and provide a simple fixed-point recursion to compute the limiting power-disparity. Fitting the model to real networks shows that minority status alone does not determine power: in gender-partitioned networks, majority homophily coupled with minority heterophily plays a central role in producing persistent disadvantage, whereas in prestige-partitioned networks a highly homophilic elite minority together with a heterophilic majority yields pronounced \emph{elitism}. Empirically, this elitism is evident when redefining “elite” from top-100 to top-50/top-20/top-10 institutions: as the elite shrinks, its relative power increases.

These results suggest that reducing disparities requires structural interventions beyond increasing representation: the model shows that simply increasing minority size need not reduce disparities, which can persist even when groups are equal in size. The most direct lever is to reduce homophily by increasing cross-group visibility and interaction (e.g., mentoring, workshops, symposia, and collaborations) and by amplifying the work of underrecognized groups. In addition, encouraging entry of new authors and ensuring they can reach established audiences can mitigate disparity, while weakening preferential attachment---for example by decoupling early-career resources and recognition from past citation counts---can reduce cumulative advantage when other factors are difficult to change. Finally, because ranking and discovery systems can amplify existing inequalities, search engines, publishers, and journals can use disparity-aware metrics to audit and diversify exposure, helping build a fairer and more inclusive web of science~\cite{linkova2020tackling}.


\bibliographystyle{ACM-Reference-Format}
\bibliography{WWW26_PowerDisparity_Refs}

\appendix
\appendix
\renewcommand\thefigure{A\arabic{figure}}
\setcounter{figure}{0}

\renewcommand\thetable{A\arabic{table}}
\setcounter{table}{0}

\vspace{0.75cm}
\noindent
{\bf Code and Dataset Availability: }All codes and complete datasets are available on GitHub to facilitate reproducibility: \\\url{https://github.com/ninoch/DMPA}

\newpage
\onecolumn
\section*{Supplementary Information}

\section{Proof of Theorem~\ref{th:convergence_DMPA}}
\label{SI:proof_of_convergence}

\subsection{Preliminaries and the outline of the proof}
The proof of Theorem~\ref{th:convergence_DMPA} uses the following result from~\cite[Theorem~5.2.1]{kushner2003}.

\begin{theorem}[Convergence w.p.~1 with Martingale difference noise~\cite{kushner2003}]
	\label{th:kushner_theorem}
	Consider the algorithm
	\begin{equation}
	\label{eq:algorithm_kushner_theorem}
	x_{\timeValue + 1} = x_{\timeValue} + \epsilon_\timeValue Y_\timeValue +\epsilon_\timeValue z_\timeValue
	\end{equation}
where,	$Y_\timeValue \in \mathbb{R}^m$, $\epsilon_\timeValue$ denotes a decreasing step size and $\epsilon_\timeValue z_\timeValue$ is the shortest Euclidean length needed to project $x_\timeValue$ into some compact set $H$. Assume,
	\begin{enumerate}[leftmargin = 0.5in]
		\item[(C.1)]  $\sup_\timeValue |Y_\timeValue|^2 < \infty$
		
		\item[(C.2)] There exists a measurable function $\bar{g}(\cdot)$ of $x$ and random variables $\beta_\timeValue$ such that, 
		\begin{equation}
		\mathbb{E}_\timeValue Y_\timeValue = \mathbb{E}\{Y_\timeValue|x_0, Y_i, i < \timeValue\} = \bar{g}(x_\timeValue) +\beta_\timeValue.
		\end{equation}Further, $Y_t = E_t Y_t + M_t$
   where $\{M_t\}$ is a bounded martingale difference process.
		
		\item[(C.3)] $\bar{g}(\cdot)$ is Lipschitz continuous
		
		\item[(C.4)] $\epsilon_\timeValue \geq 0, \epsilon_\timeValue \rightarrow 0$ for $\timeValue \geq 0$ and $\epsilon_{\timeValue} = 0$ for $\timeValue < 0$
		
		\item[(C.5)] $\sum_{\timeValue}\epsilon_{\timeValue}^2 < \infty, \sum_{i}\epsilon_i = \infty$
		
		\item[(C.6)] $\sum_{i}\epsilon_i\left|\beta_i\right| < \infty$	w.p.~1
	\end{enumerate}
If $\bar{x}$ is an asymptotically stable point of the ordinary differential equation~(ODE)
\begin{equation}
\label{eq:ODE_kushner_theorem}
 \dot{x} = \bar{g}(x) + z
\end{equation} where $z$ is the shortest Euclidean length needed to ensure $x$ is in $H$ and, $x_\timeValue$ is in some compact set in the domain of attraction of $\bar{x}$ infinitely often with probability $\geq \rho$, then $x_\timeValue \rightarrow \bar{x}$ with at least probability~$\rho$. 
\end{theorem} 

We refer the reader to \cite{kushner2003} for detailed discussion of assumptions C.1-C.6. In the context of Theorem~\ref{th:kushner_theorem}, the idea behind the proof of Theorem~\ref{th:convergence_DMPA} is to express the evolution of~$\thetaVec_\timeValue$ in the form of Eq.~\ref{eq:algorithm_kushner_theorem} with $z_\timeValue = 0$ and, show that the assumptions C.1-C.6 are satisfied. Then we show that the function~$\bar{g}$ is of the form $\bar{g}\left(\thetaVec_\timeValue\right) = \nonLinFunction(\thetaVec_\timeValue) - \thetaVec_\timeValue$ where $\nonLinFunction(\cdot)$ is a contraction map. Theorem~\ref{th:kushner_theorem} thus implies that there exists a globally asymptotically stable equilibrium state~(which is the unique fixed point of the contraction map~$\nonLinFunction(\cdot)$) to which the sequence $\{\thetaVec_\timeValue\}$ converges almost surely. 

Thus, in fact, Theorem 1 specifies more than the asymptotic normalized in- and out--degrees. It also implies that normalized in and out-degrees~(i.e.,~$\thetaVec_\timeValue$) evolving as a random process over time (appropriately scaled) converge in sample path to a deterministic ODE.

\subsection{Proof of Theorem~\ref{th:convergence_DMPA}}
{\bf Notation:} We first need some additional notation that for the proof of first part. Let,
\begin{itemize}
		\item[] $\poneBB : $ given the event~$1$ and the new node is blue, probability that it is followed by an existing blue node
		\item[] $\poneBR: $ given the event~$1$ and the new node is blue, probability that it is followed by an existing red node
		\item[] $\poneRR : $ given the event~$1$  and the new node is red, probability that  it is followed by an existing red node
		\item[] $\poneRB : $ given the event~$1$ and the new node is red, probability that it is followed by an existing blue node.
\end{itemize}
Analogous quantities for event~2 and event~3 are denoted by $\ptwoBB$, $\ptwoBR$, $\ptwoRR$, $\ptwoRB$ and $\pthreeBB$, $\pthreeBR$, $\pthreeRR$, $\pthreeRB$, respectively. Further, let $n(\calRed_\timeValue), n(\calBlue_\timeValue)$ denote the number of nodes at time $\timeValue$ in the red and blue groups respectively and $\groupSize_\timeValue
= n(\calRed_\timeValue) + n(\calBlue_\timeValue)$.

\vspace{0.2cm}
With the above notation, we can express the evolution of the total in-degree of red nodes~$\inDegree(\calRed_{\timeValue})$ as follows:
\begin{align}
\label{eq:proof_degree_i_evolution}
\mathbb{E}\{\inDegree({\calRed_{\timeValue + 1})} - \inDegree({\calRed_{\timeValue})} |
\graph_\timeValue \} &= \probEventOne(\redBirthProb\poneRR + (1-\redBirthProb)\poneBR) + \probEventTwo\redBirthProb + (1-\probEventOne - \probEventTwo)(\pthreeRR + \pthreeBR).
\end{align}
The idea behind Eq.~\ref{eq:proof_degree_i_evolution} is that the total in-degree of red group can increase by an amount of $1$ at time $\timeValue$~(i.e.,~$\inDegree({\calRed_{\timeValue + 1})} - \inDegree({\calRed_{\timeValue})} = 1$) via one of three mutually exclusive ways: event~1 takes place and either a new red node is added and follows an existing red node or a new blue node is added and follows an existing red node, event~2 takes place and a new red node is born, or event~3 takes place and an existing red node follows an existing blue or a red node. Eq.~\ref{eq:proof_degree_i_evolution} expresses the expectation of this event where the three summands correspond to the three ways in which the in-degree of the red-group increase by one. Since the number of edges in the network $\graph_\timeValue$ at time $\timeValue$ is equal to $\timeValue$~(i.e.,~$|\edgeSet_\timeValue| = \timeValue$), from Eq.~\ref{eq:proof_degree_i_evolution} we get,
\begin{align}
&\mathbb{E}\left\{\frac{\inDegree\left(\calRed_{\timeValue + 1}\right)}{\timeValue + 1}|
\graph_\timeValue \right\} =\mathbb{E}\left\{{\thetaIn_{\timeValue + 1}}|
\graph_\timeValue \right\} \nonumber\\ &\hspace{0.5cm}=\frac{\inDegree({\calRed_{\timeValue})}}{\timeValue + 1} +  \frac{1}{\timeValue + 1}\left(\probEventOne(\redBirthProb\poneRR + (1-\redBirthProb)\poneBR) + \probEventTwo\redBirthProb + (1-\probEventOne - \probEventTwo)(\pthreeRR + \pthreeBR) \right) \nonumber\\
&\hspace{0.5cm}= \thetaIn_\timeValue + \frac{1}{\timeValue + 1}\left(\probEventOne(\redBirthProb\poneRR + (1-\redBirthProb)\poneBR) + \probEventTwo\redBirthProb + (1-\probEventOne - \probEventTwo)(\pthreeRR + \pthreeBR) -\thetaIn_\timeValue\right). \label{eq:proof_theta_i_evolution}
\end{align}

Using similar arguments for the total out-degree of red group, we get,
\begin{align}
&\mathbb{E}\left\{\frac{\outDegree\left(\calRed_{\timeValue + 1}\right)}{\timeValue + 1}|
\graph_\timeValue \right\}  = \mathbb{E}\left\{{\thetaOut_{\timeValue + 1}}|
\graph_\timeValue \right\} \nonumber\\ 
&\hspace{0.5cm}= \thetaOut_\timeValue + \frac{1}{\timeValue + 1}\left( \probEventOne\redBirthProb + \probEventTwo(\redBirthProb\ptwoRR + (1-\redBirthProb)\ptwoBR) + (1-\probEventOne - \probEventTwo)(\pthreeRR + \pthreeRB) -\thetaOut_\timeValue\right). \label{eq:proof_theta_o_evolution}
\end{align}

Therefore, by combining Eq.~\ref{eq:proof_theta_i_evolution} and Eq.~\ref{eq:proof_theta_o_evolution}, we get,
\begin{align}
&\mathbb{E}\left\{{\thetaVec_{\timeValue + 1}}| \graph_\timeValue\right\}  = \thetaVec_\timeValue + \nonumber\\
&\hspace{0.25cm}\frac{1}{\timeValue + 1}
\begin{bmatrix}
\probEventOne(\redBirthProb\poneRR + (1-\redBirthProb)\poneBR) + \probEventTwo\redBirthProb + (1-\probEventOne - \probEventTwo)(\pthreeRR + \pthreeBR) -\thetaIn_\timeValue  \\
 \probEventOne\redBirthProb + \probEventTwo(\redBirthProb\ptwoRR + (1-\redBirthProb)\ptwoBR) + (1-\probEventOne - \probEventTwo)(\pthreeRR + \pthreeRB) -\thetaOut_\timeValue \label{eq:proof_theta_evolution}
\end{bmatrix}.
\end{align}

Next, let us consider~$\poneRR$. Recall that $\poneRR$ is the probability that, given the event~1 happened and the new node is red, it is followed by an existing red node. This event can happen in three ways: \\
 1~-~a potential red follower is chosen via preferential attachment~(with probability~$\frac{\inDegree(\calRed_{\timeValue}) + \groupSize(\calRed_\timeValue)\deltaboth}{\degree_\timeValue + \groupSize_\timeValue \deltaboth}$) and the new node is followed by it~(probability $\homophilyRedAnyEvent$)
  \\2~-~a potential red follower is chosen via preferential attachment~(with probability~$\frac{\inDegree(\calRed_{\timeValue}) + \groupSize(\calRed_\timeValue)\deltaboth}{\degree_\timeValue + \groupSize_\timeValue \deltaboth}$) and the new node is not followed by it~(probability $1-\homophilyRedAnyEvent$) and then the event takes place after that with probability $\poneRR$\\
 3~-~a potential blue follower is chosen via preferential attachment~(with probability~$\frac{\inDegree(\calBlue_{\timeValue}) + \groupSize(\calBlue_\timeValue)\deltaboth}{\degree_\timeValue + \groupSize_\timeValue \deltaboth}$) and the new node is not followed by it~(probability $\homophilyBlueAnyEvent$) and then the event takes place after that with probability $\poneRR$.

 Hence, $\poneRR$ satisfies,
\begin{align}
\poneRR &= \frac{\inDegree(\calRed_{\timeValue}) + \groupSize(\calRed_\timeValue)}{\degree_\timeValue + \groupSize_\timeValue \deltaboth}\homophilyRedAnyEvent + \left(\frac{\inDegree(\calBlue_{\timeValue}) + \groupSize(\calBlue_\timeValue)\deltaboth}{\degree_\timeValue + \groupSize_\timeValue \deltaboth}\homophilyBlueAnyEvent + \frac{\inDegree(\calRed_{\timeValue}) + \groupSize(\calRed_\timeValue)\deltaboth}{\degree_\timeValue + \groupSize_\timeValue \deltaboth}\left(1-\homophilyRedAnyEvent\right) \right)\poneRR \nonumber\\
 &= \frac{ \left(\inDegree(\calRed_\timeValue) + \groupSize(\calRed_{\timeValue})\deltaboth\right)\homophilyRedAnyEvent}{\degree_\timeValue + \groupSize_\timeValue \deltaboth -\left(		 \left(\inDegree(\calBlue_\timeValue) + \groupSize(\calBlue_{\timeValue})\deltaboth\right)\homophilyBlueAnyEvent	+  \left(\inDegree(\calRed_\timeValue) + \groupSize(\calRed_{\timeValue})\deltaboth\right)\left(1-\homophilyRedAnyEvent\right)\right)} \nonumber\\
 &= \frac{\left(\thetaIn_\timeValue\degree_\timeValue + \groupSize(\calRed_\timeValue)\deltaboth \right)\homophilyRedAnyEvent}{\degree_\timeValue + \groupSize_\timeValue \deltaboth -		 \left( \left(1-\thetaIn_\timeValue\right) \degree_\timeValue + \groupSize(\calBlue_{\timeValue})\deltaboth \right)\homophilyBlueAnyEvent	 -\left( \thetaIn_\timeValue \degree_\timeValue+ \groupSize(\calRed_{\timeValue})\deltaboth\right)\left(1-\homophilyRedAnyEvent\right)} \quad \text{(by definition of $\thetaIn_\timeValue$)} \nonumber\\
 &= \frac{\left(\thetaIn_\timeValue\timeValue + \groupSize(\calRed_\timeValue)\deltaboth \right)\homophilyRedAnyEvent}{\timeValue + \groupSize_\timeValue \deltaboth -		 \left( \left(1-\thetaIn_\timeValue\right) \timeValue + \groupSize(\calBlue_{\timeValue})\deltaboth \right)\homophilyBlueAnyEvent	 -\left( \thetaIn_\timeValue \timeValue+ \groupSize(\calRed_{\timeValue})\deltaboth\right)\left(1-\homophilyRedAnyEvent\right)} \quad \text{(since $\degree_\timeValue = \timeValue$)} \label{eq:poneRR}
\end{align}
Next, by Hoeffding's inequality, we get,
\begin{equation}
\label{eq:hoeffding_poneRR}
\begin{aligned}
\mathbb{P}\left\{	\left|\groupSize\left(\calRed_{\timeValue} \right) -(\probEventOne + \probEventTwo)\redBirthProb\timeValue \right|	\geq \sqrt{\frac{\timeValue \log \timeValue}{2}}\right\} &\leq \frac{1}{\timeValue}\\
\mathbb{P}\left\{	\left|\groupSize\left(\calBlue_{\timeValue} \right) -(\probEventOne + \probEventTwo)\left(1-\redBirthProb\right)\timeValue \right|	\geq \sqrt{\frac{\timeValue \log \timeValue}{2}}\right\} &\leq \frac{1}{\timeValue}\\
\mathbb{P}\left\{	\left|\groupSize_\timeValue -(\probEventOne + \probEventTwo)\timeValue \right|	\geq \sqrt{\frac{\timeValue \log \timeValue}{2}}\right\} &\leq \frac{1}{\timeValue}.
\end{aligned}
\end{equation}
Hence, by Eq.~\ref{eq:poneRR} and Eq.~\ref{eq:hoeffding_poneRR}, we observe that, with probability $1-o(\frac{1}{\timeValue})$,
\begin{align}
\poneRR&= \frac{\left(\thetaIn_\timeValue + \left(\probEventOne+\probEventTwo\right)\redBirthProb\deltaboth \right)\homophilyRedAnyEvent}{1 + \left(\probEventOne+\probEventTwo\right)\deltaboth -		 \left( \left(1-\thetaIn_\timeValue\right) + \left(\probEventOne+\probEventTwo\right)\left(1-\redBirthProb\right)\deltaboth \right)\homophilyBlueAnyEvent	 -\left( \thetaIn_\timeValue+ \left(\probEventOne+\probEventTwo\right)\redBirthProb\deltaboth\right)\left(1-\homophilyRedAnyEvent\right)} + O\left(\frac{1}{\timeValue^{\frac{1}{4}}}\right). \nonumber\\
\end{align}
Following similar steps, we also get, with probability $1-o(\frac{1}{\timeValue})$,
\begin{align}
\poneBR&= \frac{\left(\thetaIn_\timeValue + \left(\probEventOne+\probEventTwo\right)\redBirthProb\deltaboth \right)\left(1-\homophilyRedAnyEvent\right)}{1 + \left(\probEventOne+\probEventTwo\right)\deltaboth -		 \left( \left(1-\thetaIn_\timeValue\right) + \left(\probEventOne+\probEventTwo\right)\left(1-\redBirthProb\right)\deltaboth \right)\left(1-\homophilyBlueAnyEvent\right) -\left( \thetaIn_\timeValue+ \left(\probEventOne+\probEventTwo\right)\redBirthProb\deltaboth\right)\homophilyRedAnyEvent} + O\left(\frac{1}{\timeValue^{\frac{1}{4}}}\right). \nonumber\\
\ptwoRR&= \frac{\left(\thetaOut_\timeValue + \left(\probEventOne+\probEventTwo\right)\redBirthProb\deltaboth \right)\homophilyRedAnyEvent}{1 + \left(\probEventOne+\probEventTwo\right)\deltaboth -		 \left( \left(1-\thetaOut_\timeValue\right) + \left(\probEventOne+\probEventTwo\right)\left(1-\redBirthProb\right)\deltaboth \right)\homophilyRedAnyEvent	 -\left( \thetaOut_\timeValue+ \left(\probEventOne+\probEventTwo\right)\redBirthProb\deltaboth\right)\left(1-\homophilyRedAnyEvent\right)} + O\left(\frac{1}{\timeValue^{\frac{1}{4}}}\right). \nonumber\\
\ptwoBR&= \frac{\left(\thetaOut_\timeValue + \left(\probEventOne+\probEventTwo\right)\redBirthProb\deltaboth \right)\left(1-\homophilyBlueAnyEvent\right)}{1 + \left(\probEventOne+\probEventTwo\right)\deltaboth -		 \left( \left(1-\thetaOut_\timeValue\right) + \left(\probEventOne+\probEventTwo\right)\left(1-\redBirthProb\right)\deltaboth \right)\left(1-\homophilyBlueAnyEvent\right)	 -\left( \thetaOut_\timeValue+ \left(\probEventOne+\probEventTwo\right)\redBirthProb\deltaboth\right)\homophilyBlueAnyEvent} + O\left(\frac{1}{\timeValue^{\frac{1}{4}}}\right). \nonumber
\end{align} and,
\begin{align}
\pthreeRR&= \frac{\left(\thetaOut_\timeValue + \left(\probEventOne + \probEventTwo\right)\redBirthProb\deltaboth \right)\left(\thetaIn_\timeValue + \left(\probEventOne + \probEventTwo\right)\redBirthProb\deltaboth \right)\homophilyRedAnyEvent}{
	\splitfrac{\left(1+\left(\probEventOne+\probEventTwo\right)\deltaboth\right)^2 }{
	\splitfrac{
	 - \left( 1	-\thetaOut_\timeValue+ \left(\probEventOne+\probEventTwo\right)\left(1-\redBirthProb\right)\deltaboth\right)\left( 1	-\thetaIn_\timeValue+ \left(\probEventOne+\probEventTwo\right)\left(1-\redBirthProb\right)\deltaboth\right)\left(1-\homophilyBlueAnyEvent\right)}{\splitfrac{- \left( 1	-\thetaOut_\timeValue+ \left(\probEventOne+\probEventTwo\right)\left(1-\redBirthProb\right)\deltaboth\right)\left(\thetaIn_\timeValue+ \left(\probEventOne+\probEventTwo\right)\redBirthProb\deltaboth\right)\homophilyRedAnyEvent }{\splitfrac{- \left( \thetaOut_\timeValue+ \left(\probEventOne+\probEventTwo\right)\redBirthProb\deltaboth\right)\left( 1	-\thetaIn_\timeValue+ \left(\probEventOne+\probEventTwo\right)\left(1-\redBirthProb\right)\deltaboth\right)\homophilyBlueAnyEvent}{
	 	- \left(\thetaOut_\timeValue+ \left(\probEventOne+\probEventTwo\right)\redBirthProb\deltaboth\right)\left(	\thetaIn_\timeValue+ \left(\probEventOne+\probEventTwo\right)\redBirthProb\deltaboth\right)\left(1-\homophilyRedAnyEvent\right)
 	}	 
 } 
}
}
}	+ O\left(\frac{1}{\timeValue^{\frac{1}{4}}}\right)	\label{eq:pthreee_RR}\end{align}
\begin{align}
\pthreeBR&= \frac{\left(1-\thetaOut_\timeValue + \left(\probEventOne + \probEventTwo\right)\left(1-\redBirthProb\right)\deltaboth \right)\left(\thetaIn_\timeValue + \left(\probEventOne + \probEventTwo\right)\redBirthProb\deltaboth \right)\left(1-\homophilyRedAnyEvent\right)}{
	\splitfrac{\left(1+\left(\probEventOne+\probEventTwo\right)\deltaboth\right)^2 }{
		\splitfrac{
			- \left( 1	-\thetaOut_\timeValue+ \left(\probEventOne+\probEventTwo\right)\left(1-\redBirthProb\right)\deltaboth\right)\left( 1	-\thetaIn_\timeValue+ \left(\probEventOne+\probEventTwo\right)\left(1-\redBirthProb\right)\deltaboth\right)\left(1-\homophilyBlueAnyEvent\right)}{\splitfrac{- \left( 1	-\thetaOut_\timeValue+ \left(\probEventOne+\probEventTwo\right)\left(1-\redBirthProb\right)\deltaboth\right)\left(\thetaIn_\timeValue+ \left(\probEventOne+\probEventTwo\right)\redBirthProb\deltaboth\right)\homophilyRedAnyEvent }{\splitfrac{- \left( \thetaOut_\timeValue+ \left(\probEventOne+\probEventTwo\right)\redBirthProb\deltaboth\right)\left( 1	-\thetaIn_\timeValue+ \left(\probEventOne+\probEventTwo\right)\left(1-\redBirthProb\right)\deltaboth\right)\homophilyBlueAnyEvent}{
					- \left(\thetaOut_\timeValue+ \left(\probEventOne+\probEventTwo\right)\redBirthProb\deltaboth\right)\left(	\thetaIn_\timeValue+ \left(\probEventOne+\probEventTwo\right)\redBirthProb\deltaboth\right)\left(1-\homophilyRedAnyEvent\right)
				}	 
			} 
		}
	}
}	+ O\left(\frac{1}{\timeValue^{\frac{1}{4}}}\right)	\label{eq:pthreee_BR}\end{align}
\begin{align}
\pthreeRB&= \frac{\left(\thetaOut_\timeValue + \left(\probEventOne + \probEventTwo\right)\redBirthProb\deltaboth \right)\left(1-\thetaIn_\timeValue + \left(\probEventOne + \probEventTwo\right)\left(1-\redBirthProb\right)\deltaboth \right)\left(1-\homophilyBlueAnyEvent\right)}{
	\splitfrac{\left(1+\left(\probEventOne+\probEventTwo\right)\deltaboth\right)^2 }{
		\splitfrac{
			- \left( 1	-\thetaOut_\timeValue+ \left(\probEventOne+\probEventTwo\right)\left(1-\redBirthProb\right)\deltaboth\right)\left( 1	-\thetaIn_\timeValue+ \left(\probEventOne+\probEventTwo\right)\left(1-\redBirthProb\right)\deltaboth\right)\left(1-\homophilyBlueAnyEvent\right)}{\splitfrac{- \left( 1	-\thetaOut_\timeValue+ \left(\probEventOne+\probEventTwo\right)\left(1-\redBirthProb\right)\deltaboth\right)\left(\thetaIn_\timeValue+ \left(\probEventOne+\probEventTwo\right)\redBirthProb\deltaboth\right)\homophilyRedAnyEvent }{\splitfrac{- \left( \thetaOut_\timeValue+ \left(\probEventOne+\probEventTwo\right)\redBirthProb\deltaboth\right)\left( 1	-\thetaIn_\timeValue+ \left(\probEventOne+\probEventTwo\right)\left(1-\redBirthProb\right)\deltaboth\right)\homophilyBlueAnyEvent}{
					- \left(\thetaOut_\timeValue+ \left(\probEventOne+\probEventTwo\right)\redBirthProb\deltaboth\right)\left(	\thetaIn_\timeValue+ \left(\probEventOne+\probEventTwo\right)\redBirthProb\deltaboth\right)\left(1-\homophilyRedAnyEvent\right)
				}	 
			} 
		}
	}
}	+ O\left(\frac{1}{\timeValue^{\frac{1}{4}}}\right)	\label{eq:pthreee_RB}\end{align}
\begin{align}
\pthreeBB&= \frac{\left(1-\thetaOut_\timeValue + \left(\probEventOne + \probEventTwo\right)\left(1-\redBirthProb\right)\deltaboth \right)\left(1-\thetaIn_\timeValue + \left(\probEventOne + \probEventTwo\right)\left(1-\redBirthProb\right)\deltaboth \right)\homophilyBlueAnyEvent}{
	\splitfrac{\left(1+\left(\probEventOne+\probEventTwo\right)\deltaboth\right)^2 }{
		\splitfrac{
			- \left( 1	-\thetaOut_\timeValue+ \left(\probEventOne+\probEventTwo\right)\left(1-\redBirthProb\right)\deltaboth\right)\left( 1	-\thetaIn_\timeValue+ \left(\probEventOne+\probEventTwo\right)\left(1-\redBirthProb\right)\deltaboth\right)\left(1-\homophilyBlueAnyEvent\right)}{\splitfrac{- \left( 1	-\thetaOut_\timeValue+ \left(\probEventOne+\probEventTwo\right)\left(1-\redBirthProb\right)\deltaboth\right)\left(\thetaIn_\timeValue+ \left(\probEventOne+\probEventTwo\right)\redBirthProb\deltaboth\right)\homophilyRedAnyEvent }{\splitfrac{- \left( \thetaOut_\timeValue+ \left(\probEventOne+\probEventTwo\right)\redBirthProb\deltaboth\right)\left( 1	-\thetaIn_\timeValue+ \left(\probEventOne+\probEventTwo\right)\left(1-\redBirthProb\right)\deltaboth\right)\homophilyBlueAnyEvent}{
					- \left(\thetaOut_\timeValue+ \left(\probEventOne+\probEventTwo\right)\redBirthProb\deltaboth\right)\left(	\thetaIn_\timeValue+ \left(\probEventOne+\probEventTwo\right)\redBirthProb\deltaboth\right)\left(1-\homophilyRedAnyEvent\right)
				}	 
			} 
		}
	}
}	+ O\left(\frac{1}{\timeValue^{\frac{1}{4}}}\right)	\label{eq:pthreee_BB}
\end{align}

Now, we have all the ingredients to show that the assumptions C.1-C.6 are satisfied Theorem~\ref{th:kushner_theorem} for the sequence~$\{\thetaVec_\timeValue\}$. 

First, note that $\{\thetaVec_\timeValue\}$ is Markovian implying that its evolution can be expressed in the form Eq.~\ref{eq:algorithm_kushner_theorem} with $z_\timeValue = 0$ for all $\timeValue = 1, 2, \dots$ since $\thetaVec_\timeValue$ is naturally in the space $[0,1]^2$~i.e.,~$\thetaVec_{\timeValue + 1} = \thetaVec_\timeValue + \epsilon_{\timeValue}Y_\timeValue$. This further implies that C.1 and C.2 are satisfied. 

Next, note that some of the terms on the right side of Eq.~\ref{eq:proof_theta_evolution} are constant model parameters~($\redBirthProb$, $\probEventOne$, $\probEventTwo$, $\homophilyBlueAnyEvent$, $\homophilyRedAnyEvent$, $\deltaboth$) and each of the remaining terms~($\poneRR$, $\poneBR$, $\ptwoRR$, $\ptwoBR$, $\pthreeRR$, $\pthreeBR$, $\pthreeRB$) is equal to a function of $\thetaIn_\timeValue, \thetaOut_\timeValue$ plus a noise term that is $O\left(\frac{1}{\timeValue^{{1}/{4}}}\right)$ with probability $1-o\left(\frac{1}{\timeValue}\right)$. Further, we can see from Eq.~\ref{eq:proof_theta_evolution} that $\epsilon_{\timeValue} = \frac{1}{\timeValue + 1}$ and since the noise term is $O\left(\frac{1}{\timeValue^{{1}/{4}}}\right)$, C.6 is satisfied. Further, we get,
\begin{equation}
\label{eq:proof_F_expression}
\begin{aligned}
g(\thetaVec_{\timeValue})&=\nonLinFunction(\thetaVec_\timeValue) - \thetaVec_{\timeValue}\\
\nonLinFunction(\thetaVec_\timeValue) &= \begin{bmatrix}
\probEventOne(\redBirthProb\barponeRR + (1-\redBirthProb)\barponeBR) + \probEventTwo\redBirthProb + (1-\probEventOne - \probEventTwo)(\barpthreeRR + \barpthreeBR) \\
\probEventOne\redBirthProb + \probEventTwo(\redBirthProb\barptwoRR + (1-\redBirthProb)\barptwoBR) + (1-\probEventOne - \probEventTwo)(\barpthreeRR + \barpthreeRB) 
\end{bmatrix}
\end{aligned}
\end{equation}
where, $\barponeRR$, $\barponeBR$, $\barptwoRR$, $\barptwoBR$, $\barpthreeRR$, $\barpthreeBR$, $\barpthreeRB$ correspond to $\poneRR$, $\poneBR$, $\ptwoRR$, $\ptwoBR$, $\pthreeRR$, $\pthreeBR$, $\pthreeRB$, respectively, with the $O\left(\frac{1}{\timeValue^{{1}/{4}}}\right)$ noise term neglected. 

Hence, the only remaining task left is to show that $\nonLinFunction(\cdot)$ (defined in Eq.~\ref{eq:proof_F_expression}) is a contraction map. For this, after some tedious algebra, we obtain the Jacobian $J$ of $\nonLinFunction(\cdot)$ to be of a special form with respect to $\deltaboth$: each element (denoted by~$J_{ij}\, i,j = 1,2$) of $J$ is a ratio of two polynomials of $\deltaboth$ where the denominator polynomial is the higher order one. This implies that there exists $\deltaboth^{*} >0$ such that, for all $\deltaboth > \deltaboth^{*}$,
\begin{align}
\left|\left|J\right|\right|_2 \leq \left|\left|J\right|\right|_F = \left(\sum_{i}\sum_{j}\left|J_{ij}\right|^2\right)^\frac{1}{2} < 1
\label{eq:convergence} 
\end{align}
where $\left|\left|\cdot\right|\right|_F$ denotes the Frobenius norm and $\left|\left|\cdot\right|\right|_2$ denotes the  largest singular value~(spectral norm). Hence, there exists $\deltaboth^{*} >0$ such that, for all $\deltaboth > \deltaboth^{*}$, $\nonLinFunction(\cdot)$ is a contraction map.

\section{Additional Numerical Results}
\label{appn:additional_results}

\begin{figure*}
	\includegraphics[width=\textwidth,  trim={0.15cm 0.3cm 0.2cm 0.1cm},clip]{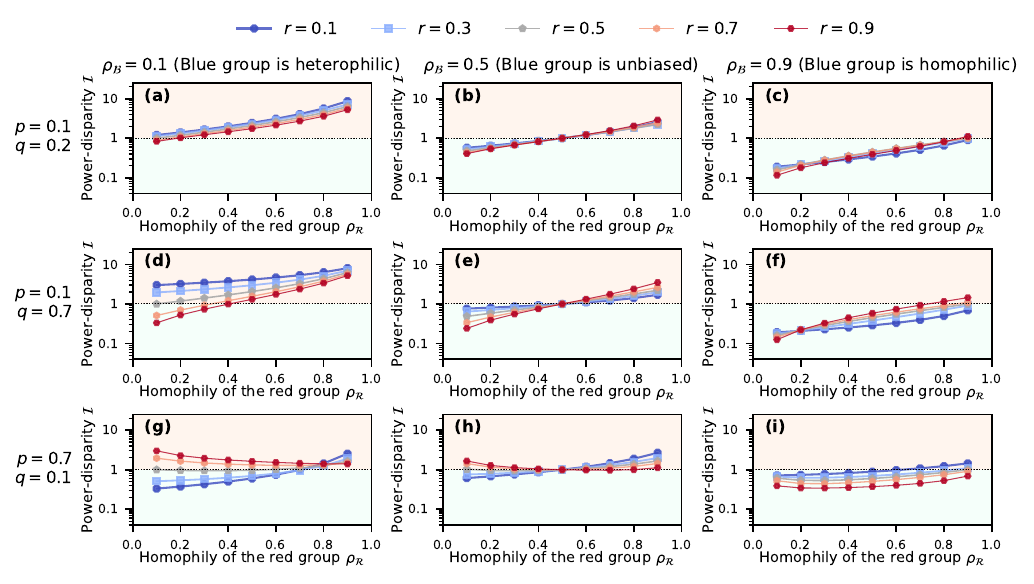}
	\caption{
The figure illustrates how preferential-attachment affects power-disparity. The plots show the variation of the asymptotic power-disparity~$\powerInequality$ with the homophily of the red group~$\homophilyRedAnyEvent$ under the DMPA model with $\deltaboth = 100$ analogous to Fig.~\ref{fig:PowerInequalityTheoretical} (where $\deltaboth = 10$).
Comparing this figure with Fig.~\ref{fig:PowerInequalityTheoretical} shows that increasing $\deltaboth$~(i.e.,~reducing the preferential attachment) reduces the effect of the group size on power-disparity, especially when one group is highly homophilic and the other heterophilic. Thus, reducing preferential-attachment is an effective strategy to ameliorate power-disparity in the presence of extreme class-imbalance and differences in homophily parameters of the two groups.
} 
	\label{fig:PowerInequalityTheoretical_delta100}
\end{figure*}

\paragraph{Power-disparity values for $\deltaboth = 100$~(reduced preferential attachment):} Fig.~\ref{fig:PowerInequalityTheoretical_delta100} illustrates the asymptotic power-disparity $\powerInequality$ under the DMPA model for different parameter combinations with fixed $\deltaboth = 100$. Comparing Fig.~\ref{fig:PowerInequalityTheoretical}~(generated with $\deltaboth = 10$) with Fig.~\ref{fig:PowerInequalityTheoretical_delta100} shows that increasing $\deltaboth$ (i.e.,~reducing preferential-attachment) reduces the power-disparity. In particular, the effect of $\deltaboth$ on power-disparity is larger when there is a high degree of class imbalance (i.e.,~$\redBirthProb \ll 0.5$ or $\redBirthProb \gg 0.5$).

\begin{figure}
\centering
	\includegraphics[width=0.75\columnwidth,  trim={0.15cm 0.2cm 0.2cm 0.2cm},clip]{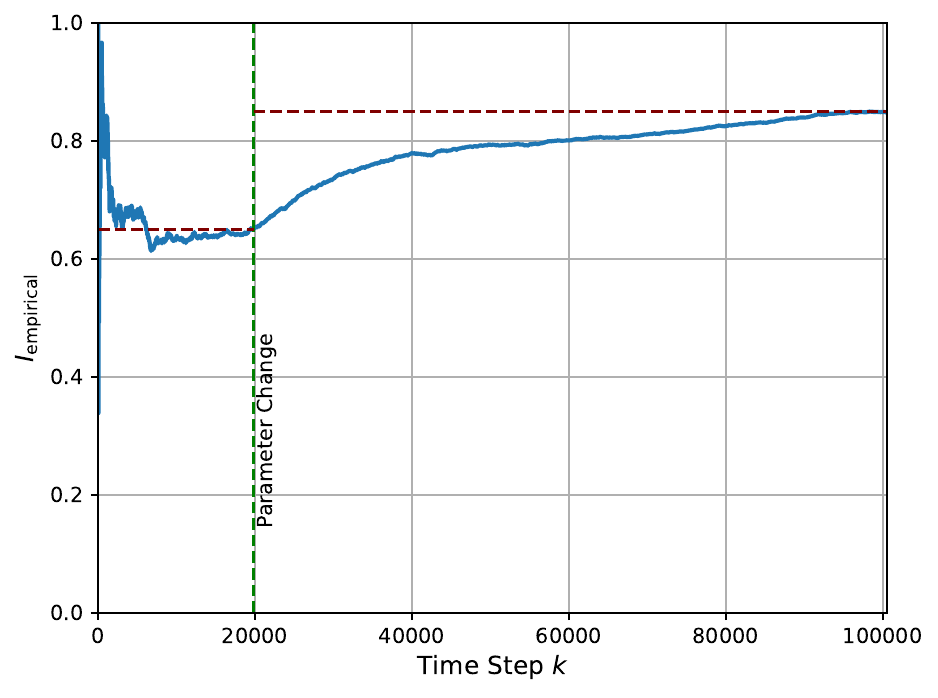}
    \vspace{-0.2cm} 
	\caption{ The evolution of the power-disparity $\powerInequality_\timeValue$ resulting from the proposed model before and after a change in the field. The field before the parameter change corresponds to a ``bad" scenario with fewer women being recruited ($\redBirthProb = 0.2$)  homophilic males and  heterophilic females ($\homophilyRedAnyEvent = 0.4, \homophilyBlueAnyEvent = 0.6$) and $p = 0.01, q = 0.09, \delta = 10$. After $20,000$ steps, the field has become ``better" with more women being recruited ($\redBirthProb = 0.4$) compared to before and both groups being more neutral ($\homophilyRedAnyEvent = 0.45, \homophilyBlueAnyEvent = 0.55$). The evolution of the empirical power-disparity at each time instant $\powerInequality_\timeValue$~(the blue continuous line) and the theoretically predicted power-disparity~$\powerInequality$~(red dashed-line) are indicated in Fig.~\ref{fig:ParameterChange}. This experiment illustrates that the model and the definition can be useful to understand the dynamics even when parameters change.
	} 
	\label{fig:ParameterChange}
    \vspace{-0.45cm}
\end{figure}

\paragraph{A field with changing dynamics:} The proposed model and the definition of power disparity can be used to analyze a field with changing dynamics as well. For example, let us consider an academic field with fewer female researchers~($r = 0.2$) and high homophily/heterophily disparity~($\homophilyRedAnyEvent = 0.4, \homophilyBlueAnyEvent = 0.6$). In this state, the model predicts that the power-disparity $\powerInequality_\timeValue$ will become severe over time with female researchers ultimately facing an unfair disparity. In other words, $\powerInequality_\timeValue$ will gradually reach and stay at a value smaller than $1$ as per Theorem~\ref{th:convergence_DMPA}. This can be seen from the results of the simulation experiment presented in Fig.~\ref{fig:ParameterChange}: the empirical power-disparity~(blue line) gradually approaches the theoretical power-disparity~(0.65) indicated by the red line. Let us now assume that the field in consideration implemented a set of positive changes~(for example some of the mitigation strategies that we have proposed) such as making the field more appealing to the early career researchers, the homophily levels less biased~($\homophilyRedAnyEvent = 0.45, \homophilyBlueAnyEvent = 0.55$) and increased recruitment of female authors ($\redBirthProb \approx 0.4$). If we turn to the model to evaluate the effects of these implemented changes on the power-disparity, it predicts that the disparity will gradually become better~i.e.,~$\powerInequality_\timeValue$ will gradually reach a value closer to $\approx 1$ over time. More specifically, the 
Theorem~\ref{th:convergence_DMPA} implies that irrespective of the previous ``bad" state of the field, power-disparity of the field $\powerInequality_\timeValue$  will move gradually towards a value more closer to $1$ and will become stationary at that value. It can be seen from Fig.~\ref{fig:ParameterChange} that the empirical power disparity moves closer to the theoretically predicted power-disparity after the changes. As the positive changes implemented by the field start to gradually take effect, the power-disparity of the field will become steady at the value predicted by the model. Thus, the theoretical analysis of the model combined with the empirical insights help identify the most efficient changes to be made to move the field in the right direction based on its current state. The proposed disparity definition can be used to monitor the effects of the implemented changes in a data-driven manner to understand whether the field is heading in the intended direction.

\begin{table}[htbp]
\centering
\caption{Edge Statistics of Gender-Partitioned and Affiliation-Partitioned Networks Shown in Fig.~\ref{fig:power}(a) and Fig.~\ref{fig:power}(b)}
\label{tab:edges}
\begin{tabular}{|c|c|c|}
\hline
\multirow{2}{*}{\textbf{Edge Type}} & \multicolumn{2}{c|}{\textbf{Number of Edges}} \\ \cline{2-3} 
                                     & \textbf{Gender-Partnd.} & \textbf{Affiliation-Partnd.} \\ \hline
Majority to Majority                 & 801                           & 324                                \\ \hline
Majority to Minority                 & 161                           & 488                                \\ \hline
Minority to Minority                 & 112                           & 265                                \\ \hline
Minority to Majority                 & 300                           & 79                                 \\ \hline
\end{tabular}
\end{table}

\begin{table}[htbp]
\centering
\caption{Node Statistics of Gender-Partitioned and Affiliation-Partitioned Networks Shown in Fig.~\ref{fig:power}(a) and Fig.~\ref{fig:power}(b)}
\label{tab:nodes}
\begin{tabular}{|c|c|c|}
\hline
\textbf{Node Type} & \textbf{Gender-Partitioned} & \textbf{Affiliation-Partitioned} \\ \hline
Minority            & 72                           & 86                               \\ \hline
Majority            & 138                          & 185                              \\ \hline
\end{tabular}
\end{table}

\paragraph{Edge and node-level statistics of subgraphs shown in Fig.~\ref{fig:power}(a) and Fig.~\ref{fig:power}(b)}: Table~\ref{tab:edges} and Table~\ref{tab:nodes} show the edge and node level statistics of the subgraphs shown in Fig.~\ref{fig:power}(a) and Fig.~\ref{fig:power}(b). It can be seen that the majority in gender-partitioned networks~(i.e.,~male authors) receive more attention from both minority~(300 out of 412 citations) and majority~(801 out of 962 citations) groups compared to the relative size of their group~(138 out of 210) while the opposite is true for the female group. The opposite of this observation holds for the affiliation-partitioned networks: the minority~(i.e.,~authors affiliated with top-ranked institutes) receive more attention from both minority and majority groups compared to the size of their group.

\section{Additional Details on Empirical Results}
\label{appn:empirical_details}
\subsection{Data}
\definecolor{Grey}{gray}{0.9}
\begin{table}[tb]
    \centering
    \begin{tabular}{|c||c|c|c||c|c|c|}
        \hline 
        \multirow{2}{*}{\textit{Field of study}} &        
        \multicolumn{3}{c||}{\cellcolor{Grey}  \textit{Gender-partitioned Network}} & \multicolumn{3}{c|}{\cellcolor{Grey}  \textit{Affiliation-partitioned Network}} \\
         \cline{2-7}
            & \# Nodes & Density & \% Female & \# Nodes & Density & \% Elite \\
         \hline 
         Management & $111K$ & {9.9$\times 10^{-5}$} & $35.19$ & $53K$ & {3.0$\times 10^{-4}$} & $12.19$ \\ 
         \hline 
         Physics & $164K$ & {6.4$\times 10^{-4}$}& $16.39$ & $156K$ & {6.3$\times 10^{-4}$} & $21.59$ \\
         \hline 
         Psychology & $873K$ & {1.7$\times 10^{-5}$} & $49.65$ & $667K$ & {2.9$\times 10^{-5}$} & $23.18$ \\ 
         \hline 
         Political Science & $1.03M$ & {7.0$\times 10^{-6}$} & $34.20$ & $204K$ & {6.4$\times 10^{-5}$} & $17.44$ \\ 
         \hline 
         Economics & $1.3M$ & {5.6$\times 10^{-5}$} & $28.01$ & $723K$ & {1.8$\times 10^{-4}$} & $13.00$ \\ 
         \hline 
         Computer Science & $7.62M$ & {7.5$\times 10^{-6}$} & $25.79$ & $3.27M$ & {3.9$\times 10^{-5}$} & $9.31$\\ 
         \hline 
    \end{tabular}
    \vspace{0.1in}
    \caption{Information about the data. Data for all fields of study, except Physics, came from Microsoft Academic Graph, and Physics data was provided by the American Physical Society. The number of authors with known gender is larger than number of authors with known affiliation. The affiliation network has higher density, potentially confounded by the fact that authors with known affiliation are more active in publishing and citing other authors. 
    }
    \label{tab:data_info}
\end{table}

We use bibliographic data from Microsoft Academic Graph (MAG) and American Physical Society (APS). The APS data was used to analyze the citations in the field of
Physics, and MAG was used to study the remaining fields, namely, Management, Psychology, Political Science, Economics and Computer Science. In this section, we discuss potential biases introduced by our data collection and processing methods as well as their potential impact on our analysis.

To extract an author's gender, we use the Gender-API (\url{https://www.gender-api.com/}). This state-of-the-art API infers the gender based on the author's name. However, it fails to recognize the gender of authors who use initials instead of their full first names (e.g. ``J. Doe'' instead of ``John Doe'') and some Chinese and Korean names.  
Figure~\ref{fig:data_coverage} illustrates the temporal variation of the coverage for gender (i.e., the fraction of authors whose gender was inferred with Gender-API). For all fields of study except Computer Science, the coverage exceeds 85\% and remains approximately constant over time. For Computer Science, coverage starts from approximately 65\% in 1990 and gradually reaches 80\% in 2018.
We removed all authors whose gender was unknown.
Additional robustness checks to validate the above approach are given in Sec.~\ref{SI:sec:robustness_checks}.

In addition to gender, we also analyze the prestige of the author's affiliation. MAG extracts author affiliations from publications and links it to a unique ID. Due to the challenges of automatically extracting affiliations from publications, the affiliation coverage for academic fields of study in the MAG data is relatively low as seen from Fig.~\ref{fig:data_coverage}.  In contrast, affiliations in the APS data are specified by authors and stored as strings. We use string processing and normalization to map the affiliation to a unique name, which is then mapped to the ranking. Specifically, taking the full address of author's institutional affiliation, we remove stop words (e.g., ``the'') and convert all characters to lower-case. After separating the country, city, institution and department names of the affiliation (using python packages such as geonamescache), we string matched words \textit{``university''}, \textit{``institute''}, \textit{``laboratory''} or \textit{``college''} in different languages (e.g., \textit{``università''}, \textit{``universidad''}, \textit{``institut''}). We then matched the extracted and normalized institution names to those listed in Shanghai University Ranking. The coverage of affiliation for the APS data (Physics) is more than $97\%$. 
We consider authors without affiliation as not being affiliated with top-ranked institutions. 
While this could potentially bias the data (for example, when top-ranked authors' affiliation is not known), Fig.~\ref{fig:elite_range} provides evidence that this is not the case. The monotonicity of the trend for power-disparity suggests that the affiliations data is not systematically biased.

\begin{figure}[h!]
    \centering
    \includegraphics[width=0.7\columnwidth]{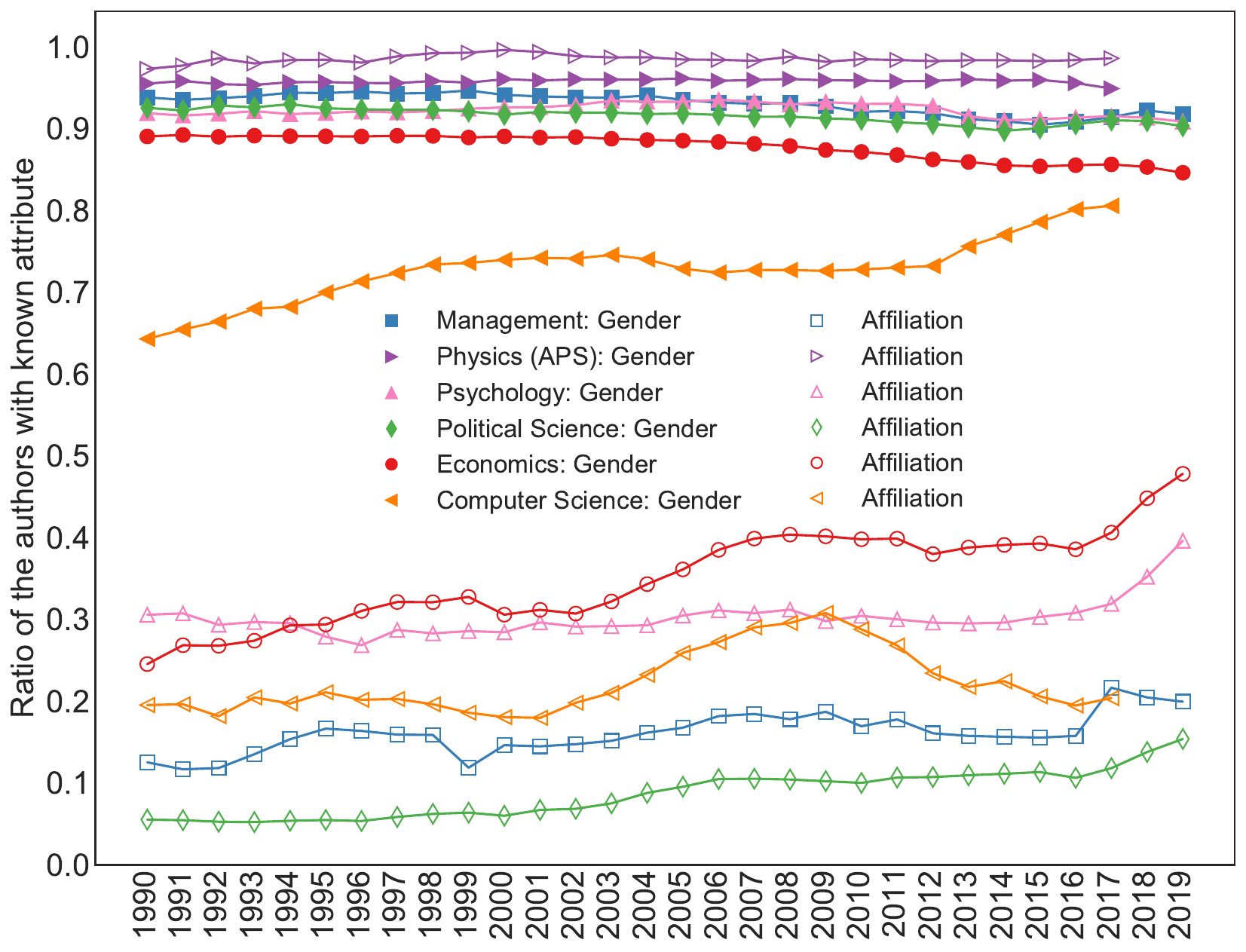}
    \caption{The fraction of authors with a known attribute (i.e.,~the coverage) over the years. The gender attribute has better coverage compared to affiliation of authors. We discard authors with unknown gender, and we consider authors with unknown affiliation in majority group (not affiliated with top-ranked universities).  
    Several sanity checks were used to verify that parameter $\redBirthProb$ estimated using this approach (for both gender and affiliation) is correct as discussed in Appendix~\ref{SI:sec:robustness_checks}.
    }
    \label{fig:data_coverage}
\end{figure}
        

\subsection{Estimating model parameters from data}
\label{SI:sec:parameter_estimation}
	{\bf Overview of the estimation method:} Our bibliometric data usually includes only the year of publication, and as a result, we do not know the order in which citation edges appear at a finer temporal resolution. To better match the conditions of the DMPA model, we create an ordered list of edges $E_{ord}$ as follows (detailed subsequently in this section). We initialize a set of nodes $S$ with one of the nodes from the largest connected component of the citation graph for 1990. Then in each iteration of the algorithm (for each year), we shuffle all the edges that have at least one end in set $S$, and append all of them to the end of list $E_{ord}$. Then we update set $S$ with the nodes covered by this newly added batch of edges. We continue this process until there are no more edges from that year that are connected to $S$. We then drop all the remaining edges from that year that are not part of the connected component and continue the procedure for the following year. 
	
 We use the nodes and edges in the ordered edge list to estimate all parameters of the DMPA model directly from the data, except for the preferential attachment parameter $\deltaboth$. We estimate the preferential attachment parameter through hyper-parameter tuning. Specifically, we use several different values of $\deltaboth$ to estimate the remaining parameters of the DMPA model, and use them to calculate the theoretical value of power-disparity. We then choose the value of $\deltaboth$ that leads to power-disparity closest to its empirical value. The parameter estimation procedure is not sensitive to the choice of the initial node to populate the seed set, nor the order in which edges are added. Repeating the parameter estimation procedure multiple times yields parameter estimates with standard deviation less than $10^{-4}$.

Table~\ref{tab:params2} provides the parameters estimated using the above outlined method for the affiliation-partitioned networks. Table~\ref{tab:params2} supplements the Table~\ref{tab:params}(estimated parameters for the gender-partitioned networks) in the main text. 

\FloatBarrier
\definecolor{Grey}{gray}{0.9}
\begin{table}[htbp]
    \centering
    \begin{tabular}{|c|c|c|c|c|c|c|}
    \hline
        \rowcolor{Grey}
        \multicolumn{7}{|c|}{\textit{Affiliation-partitioned Network Parameters}} \\
        \hline 
        \rowcolor{Grey}
        	& \textit{Mgmt.} & \textit{Phys.} & \textit{Psych.} & \textit{Poli. Sci.} & \textit{Econ.} & \textit{Com. Sci.} \\
        \hline 
        $T$ & $842.85K$ & $15.25M$ & $12.61M$ & $2.61M$ & $91.51M$ & $419.24M$ \\ 
        \hline 
        $\deltaboth$ & $1000$ & $1000$ & $1000$ & $10$ & $1000$ & $1000$ \\ 
        \hline 
        $r$ & $0.12$ & $0.22$ & $0.24$ & $0.18$ & $0.13$ & $0.09$ \\ 
        \hline 
        $p$ & $0.012$ & $0.001$ & $0.024$ & $0.020$ & $0.001$ & $0.001$\\ 
        \hline 
        $q$ & $0.048$ & $0.009$ & $0.026$ & $0.053$ & $0.006$ & $0.007$\\ 
        \hline 
        $\homophilyRedAnyEvent$ & $0.83$ & $0.72$ & $0.74$ & $0.72$ & $0.81$ & $0.85$ \\
        \hline 
        $\homophilyBlueAnyEvent$ & $0.29$ & $0.44$ & $0.47$ & $0.44$ & $0.30$ & $0.31$ \\
        \hline 
        $\powerInequality_{\mathrm{emp.}}$ & $3.11$ & $1.52$ & $1.11$ & $1.86$ & $2.31$ & $1.91$ \\
        \specialrule{.12em}{.1em}{.1em}
        $\powerInequality_{\mathrm{thr.}}$  & $3.53$ & $1.94$ & $1.79$ & $3.32$ & $3.89$ & $4.54$ \\ 
        \hline 
    \end{tabular}
    \vspace{0.1in}
    \caption{Estimated parameters of the DMPA model for the affiliation-partitioned networks. These values are also illustrated visually in Fig.~\ref{fig:estimated_parameters} (alongside corresponding values for gender-partitioned networks). Details on the datasets and estimation process are given in Appendix~\ref{appn:empirical_details}.}
    \label{tab:params2}
\vspace{-0.7cm}
\end{table}

\subsubsection{Citation Edge Ordering}
In our proposed DMPA model, citation edges form asynchronously and at a finer temporal scale than supported by the empirical data. To better align the model with the data, 
we assign an ordering to the edges so as to keep track of the new edges formed in the three events (citations to or from a new node, or between existing nodes). We create an ordered list of edges $E_{ord} = [(u_1, v_1), (u_2, v_2), ..., (u_n, v_n)]$ as follows. Starting with a \textit{seed} set with one author from the largest connected component of authors publishing in 1990, we traverse citation edges originating or terminating at these nodes and add them to the edge list $E_{ord}$ in random order. We then update the seed set by adding nodes that originate or terminate at the seed nodes and repeat the procedure until we cover all the years. The  algorithm is specified in more detail below: 

\vspace{0.2cm}
\begin{algorithm}[H]
\SetAlgoLined

\SetKwInOut{Input}{input}
\SetKwInOut{Output}{output}
\Input{graph $G$}
\Output{list of edge-ordering $E_{ord}$}

 construct graph $H$ using edges of $G$ in year $1990$\;
 
 pick one of the authors from the largest connected component of $H$ and name it as $v$\; 
 
 initialize seed $S = \{ v \}$ and $E_{ord} = []$\;
 
 \For{$y$ from $1990$ to $2020$}{
    \While{There is any potential edges to add from year $y$} {
        define $A$ as a list of edges which at least one of its sides are inside the seed set $S$\;
        
        shuffle list $A$\;
        
        append elements of $A$ to the end of $E_{ord}$\;
        
        update seed $S$ if there is a new node covered by $E_{ord}$\;
    }
 }
 \caption{Ordering of citation edges}
 \label{alg:ordering_citations}
\end{algorithm}
\vspace{0.2cm}

\noindent 
Note that the Algorithm~\ref{alg:ordering_citations} preserves the semi-dynamic ordering, meaning that there is no edge $a$ forming before edge $b$ in $E_{ord}$ where $a$ was created after $b$. Some of the edges in the citation graph $G$ will not appear in the edge list $E_{ord}$. However, since these edges are not part of the largest connected component of $G$, they likely connect to authors outside the field of study. 
Using this algorithm we cover at least $97\%$ of the edges forming during the period 1990--2020 in each field of study. We ran the parameter estimation for five times, and the standard deviation of the estimated parameters was less than $\num{1e-4}$ for all the parameters. This suggests that the randomization of the edges does not change the estimated parameters.

Next, we estimate model parameters using the new edge ordering. The number of edges generated by the arrival of new nodes (the first two types of citation events) is small compared to the number of edges generated between the existing nodes~(densification events) and therefore, empirical estimation of $\matrixEventOne$ and $\matrixEventTwo$ is not accurate due to lack of data points. 
As a result, we focus on densification events, where edges form between existing nodes, to estimate $\matrixEventThree$. We then assume that  $\matrixEventOne = \matrixEventTwo = \matrixEventThree = \matrixAnyEvent$ similar to Theorem~\ref{th:convergence_DMPA}. First, we estimated parameters $\redBirthProb$, $\probEventOne$, $\probEventTwo$, $\thetaIn$ and $\thetaOut$ using data as discussed in detail below. We then used a hyper-parameter tuning technique to estimate the preferential attachment parameter~$\deltaboth$. 
Finally, we used all the estimated parameters discussed so far in order to estimate
$\homophilyRedAnyEvent$ and $\homophilyBlueAnyEvent$ which constitute the elements of the matrix
$\matrixAnyEvent$.

\subsubsection{Estimating Class Balance Parameter $\redBirthProb$}
Recall that $\redBirthProb \in [0,1]$ represents the fraction of red nodes. 
We label authors by their gender or prestige of their institutional affiliation as described in the Methods section. For gender-partitioned networks, $\redBirthProb$ is the fraction of authors with female or unisex names. Note that this may overestimate the fraction of female authors. For affiliation-partitioned networks, $\redBirthProb$ is the fraction of authors from top-ranked institutions as defined by the Shanghai University Ranking.

\subsubsection{Estimating Edge Formation Rates $\probEventOne$, $\probEventTwo$}
For estimating the parameters $\probEventOne$ and $\probEventTwo$, we need to keep track of newly joined nodes over time. So, starting from empty set $S$, we iterate over all the edges in $E_{ord}$ and for each edge $(u, v)$ we add both $u$ and $v$ to set $S$ if they are not in the set already. Having $S$, parameter $\probEventOne$ (resp. $\probEventTwo$) could be estimated by counting the number of outgoing edges from (resp. incoming edges to) the newly added nodes to set $S$. In other words, we need to count the number of times $u$ (cited node) was not in set $S$ and the number of times $v$ (citing node) was not in the set to estimate $p$ and $q$ respectively. 

\subsubsection{Estimating Preferential Attachment Parameter $\deltaboth$}
We find the best value of $\deltaboth$ (i.e.,~the value that results in the best set of model parameters in the sense of maximum likelihood)  through a hyper-parameter tuning approach as follows. We first estimate all parameters using different values of $\deltaboth \in \{1, 2, 3, 4, 5, 10, 20, 50, 100, 1000\}$ and use them to calculate the theoretical value of power-disparity $\powerInequality_{theoretical}$, if the condition given in  Eq.~\ref{eq:convergence}~(to ensure the convergence) is satisfied. We then select the parameter set that leads to power-disparity value $\powerInequality$ that is closest to its empirical value. 

\subsubsection{Estimating Homophily Parameters $\homophilyRedAnyEvent$, $\homophilyBlueAnyEvent$}
We estimate $\homophilyRedAnyEvent$ and $\homophilyBlueAnyEvent$ using the number of generated edges. Considering only the edges forming between existing nodes, we define $N_{rr}$ as the number of edges where both 
citing and cited nodes are red nodes, and $N_{xr}$ as number of edges where the citing node is red and the cited node could be red or blue. Then we can calculate $h_{\mathcal{R}} = \frac{N_{rr}}{N_{xr}}$ from the data. $\pthreeRR$ (given in Eq.~\ref{eq:pthreee_RR}) and $\pthreeBR$ (given in Eq.~\ref{eq:pthreee_BR}) are the probabilities of generating a red-to-red and blue-to-red edges in event 3, respectively. Using those expressions, we can write $h_\mathcal{R}$ as:

\begin{equation}
\frac{N_{rr}}{N_{xr}} = h_{\mathcal{R}} = \frac{\pthreeRR}{\pthreeRR + \pthreeBR} = \frac{a \times \homophilyRedAnyEvent}{a \times \homophilyRedAnyEvent + b \times (1 - \homophilyRedAnyEvent)}
\rightarrow \homophilyRedAnyEvent = \frac{b \times h_{\mathcal{R}}}{a - a \times h_{\mathcal{R}} + b \times h_{\mathcal{R}}}
\label{eq:rur_esimate}
\end{equation}

Here $a$ and $b$ are: 
\begin{equation}
a = \left(\thetaOut_\timeValue + \left(\probEventOne + \probEventTwo\right)\redBirthProb\deltaboth \right)\left(\thetaIn_\timeValue + \left(\probEventOne + \probEventTwo\right)\redBirthProb\deltaboth \right)
\end{equation}

\begin{equation}
b = \left(1-\thetaOut_\timeValue + \left(\probEventOne + \probEventTwo\right)\left(1-\redBirthProb\right)\deltaboth \right)\left(\thetaIn_\timeValue + \left(\probEventOne + \probEventTwo\right)\redBirthProb\deltaboth \right)
\end{equation}

Similarly, based on Eq.~\ref{eq:pthreee_BB} and Eq.~\ref{eq:pthreee_RB}, we have: 

\begin{equation}
\frac{N_{bb}}{N_{xb}} = h_{\mathcal{B}} = \frac{\pthreeBB}{\pthreeBB + \pthreeRB} = \frac{c \times \homophilyBlueAnyEvent}{c \times \homophilyBlueAnyEvent + d \times (1 - \homophilyBlueAnyEvent)}
\rightarrow \homophilyBlueAnyEvent = \frac{d \times h_{\mathcal{B}}}{c - c \times h_{\mathcal{B}}  + d \times h_{\mathcal{B}} },
\label{eq:rub_esimate}
\end{equation}

\noindent where $c$ and $d$ are

\begin{equation}
c = \left(1 - \thetaOut_\timeValue + \left(\probEventOne + \probEventTwo\right)\left(1-\redBirthProb\right)\deltaboth \right)\left(1-\thetaIn_\timeValue + \left(\probEventOne + \probEventTwo\right)\left(1-\redBirthProb\right)\deltaboth \right)
\end{equation}

\begin{equation}
d = \left(\thetaOut_\timeValue + \left(\probEventOne + \probEventTwo\right)\redBirthProb\deltaboth \right)\left(1-\thetaIn_\timeValue + \left(\probEventOne + \probEventTwo\right)\left(1-\redBirthProb\right)\deltaboth \right)
\end{equation}

\noindent Parameter $\thetaIn$ is the fraction of the total in-degree of red group and $\thetaOut$ is fraction of the total out-degree of red group, and could be estimated from data. Parameters $\redBirthProb$, $\probEventOne$ and $\probEventTwo$ and $\deltaboth$ could be estimated as described above. So, we can estimate $\homophilyRedAnyEvent$ and $\homophilyBlueAnyEvent$ using other parameters and equations \ref{eq:rur_esimate} and \ref{eq:rub_esimate}. 
Note that we are assuming citation networks have fixed parameters over time (i.e. $\thetaIn$, $\thetaOut$, $\redBirthProb$, ...). However, this is not the case in the citation network. The ratio of female/elite authors, for example, changes over time. This could explain the disparity between theoretical power-disparity and empirical power-disparity in Table~\ref{tab:params}.

\section{Robustness Checks for Empirical Results on Gender Disparities}
\label{SI:sec:robustness_checks}

\subsection{A Model-Agnostic Test for the Homophily Levels}

\begin{table}[htbp]
\centering
\begin{tabular}{l|rrr|rrr}
\toprule
{} &  $n_{m}$ &  $h_{mm}$ &  \makecell{p-value\\($H0: h_{mm}< n_{m}$)} &  $n_{f}$&  $h_{ff}$ &  \makecell{p-value\\($H0: h_{ff}> n_{f}$)} \\
\midrule
Management        &            0.650 &                             0.749 &                                       0.000 &              0.350 &                                 0.319 &                                         0.000 \\
Physics               &            0.857 &                             0.904 &                                       0.000 &              0.143 &                                 0.134 &                                         0.001 \\
Psychology        &            0.549 &                             0.665 &                                       0.000 &              0.451 &                                 0.455 &                                         0.950 \\
Political Science &            0.659 &                             0.792 &                                       0.000 &              0.341 &                                 0.320 &                                         0.000 \\
Economics         &            0.736 &                             0.840 &                                       0.000 &              0.264 &                                 0.220 &                                         0.000 \\
Computer Science                &            0.838 &                             0.871 &                                       0.000 &              0.162 &                                 0.191 &                                         1.000 \\
\bottomrule
\end{tabular}
\vspace{0.1cm}
\caption{The table summarizes the results of the hypothesis tests for the homophily of the male group and the heterophily of the female group. These model-agnostic tests show the exact same observations drawn from the estimated homophily parameters of the model: male group is homophilic in all six fields~(p-value $< 0.01$) and the female group is heterophilic in all fields except Computer Science and Psychology~(p-value $< 0.01$).}
\label{tab:ComparingHomophily}
\end{table}

\begin{figure}
\centering
	\includegraphics[width=0.75\columnwidth,  trim={0.15cm 0.15cm 0.2cm 0.2cm},clip]{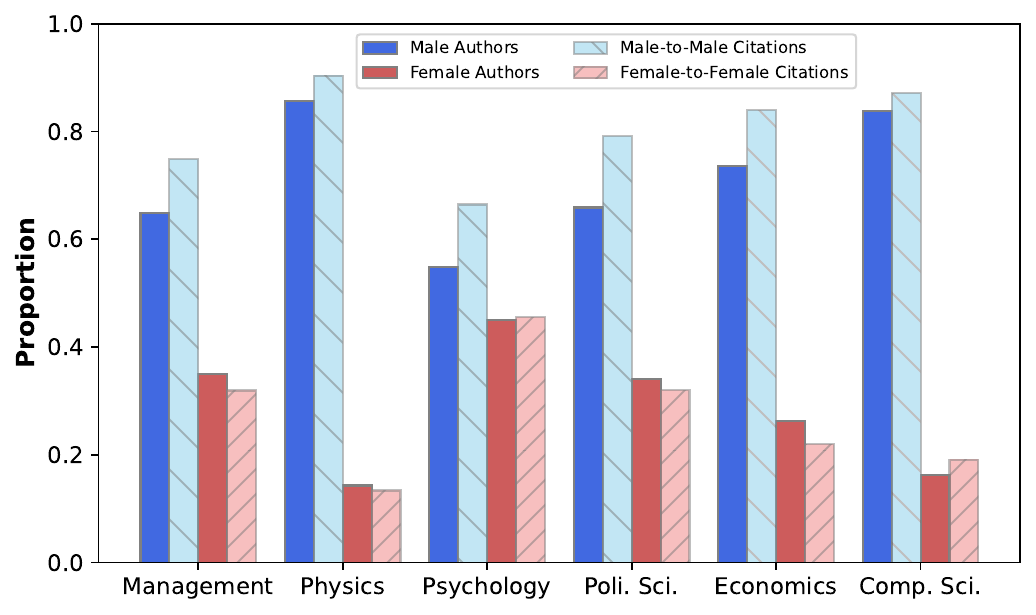}
	\caption{A visualization of the model-agnostic quantities summarized in Table~\ref{tab:ComparingHomophily}. It can be seen that the citations from the male group are more likely to go to other male authors, indicating the consistent homophily of the male group. On the other hand, citations from female authors are more likely to go to male authors in all fields except Psychology and Computer Science. These model-agnostic observations are statistically significant as seen from the p-values in Table~\ref{tab:ComparingHomophily} and they agree exactly with the model-based observations related to the homophily levels of the two groups.
	} 
	\label{fig:ComparingHomophily}
\end{figure}

We recall that the estimated values of the model parameters using gender-partitioned networks showed that the male group is highly homophilic in each field~($\homophilyBlueAnyEvent>0.5$, i.e., they are likely to cite other male authors). Additionally, the female group is heterophilic~($\homophilyRedAnyEvent<0.5$, i.e., they are unlikely to cite other female authors) in each field except computer science and psychology where the female group is also homophilic~($\homophilyRedAnyEvent>0.5$). This can be seen from Table~\ref{tab:params} and Fig.~\ref{fig:estimated_parameters}. The proposed model explains how the high homophily of the male group and the heterophily of the female group is a key factor in high-levels of power-disparity observed in Economics, Political Science, Management and Physics. The analysis~(Fig.~\ref{fig:PowerInequalityTheoretical}) also suggests that eliminating the observed homophily and heterophily from both groups is the most efficient way to mitigate the power-disparity.

To further justify above claims that are based on the homophily and heterophily of the male and female authors, we perform hypothesis testing based on a model-agnostic measure of homophily level for each group in each field. Specifically, let $h_{mm}$ be the number of citations made by male authors to other male authors as a fraction of all citations made by male authors, and $n_{m}$ be the fraction of male authors. Let $h_{ff}$ and $n_{f}$ be the analogous quantities for female authors. In a random sample of citations, we expect $h_{mm}$ to be approximately equal to $n_{m}$ if the male authors are unbiased. If $h_{mm}>n_{m}$, it would suggest that the male group is homophilic since the number of citations they make to their own group is larger than
expected under unbiasedness. Similar statements hold for female group as well. Table~\ref{tab:ComparingHomophily} presents the results of hypothesis test~(one-sided z-test for the difference in population proportions) for the homophily of the male group~(null hypothesis $H0: h_{mm}<n_{m}$) and the heterophily of the female group~(null hypothesis $H0: h_{ff}>n_{f}$) on a sample of 100,000 random citations. These rates are also shown in Fig.~\ref{fig:ComparingHomophily}. 

Several observations can be drawn from this experiment:
\begin{itemize}
    \item Firstly, the test results in Table~\ref{tab:ComparingHomophily} verify that the male group is homophilic in each field (p-value $<0.01$) and female group is heterophilic in all fields except computer science and psychology (p-value $<0.01$). Visually, this can be observed from Fig.~\ref{fig:ComparingHomophily} where the male group makes disproportionately more citations to their group in each field. This model-agnostic result agrees precisely with the model-based observations related to homophily levels obtained in our empirical analysis as we emphasized before~(based on Table~\ref{tab:params} and supplementary Fig.~\ref{fig:estimated_parameters}).

    \item Secondly, for psychology and computer science, the female group is homophilic as seen from the high p-values in Table~\ref{tab:ComparingHomophily} for the corresponding rows~(we also performed the test with the null hypothesis being that the female group is homophilic and it yielded the same conclusion). 
\end{itemize}

Thus, these additional hypothesis tests using a model-agnostic metric of the homophily confirm the observations drawn from the empirically estimated homophily parameters of the model~($\homophilyRedAnyEvent, \homophilyBlueAnyEvent$). This strengthens the homophily aspect of our model and highlights its centrality to the proposed mitigation strategies.

\subsection{Robustness of the Calculated Power-Disparity Values}

    \begin{figure}
    \centering
	\includegraphics[width=0.75\columnwidth,  trim={0.15cm 0.15cm 0.2cm 0.2cm},clip]{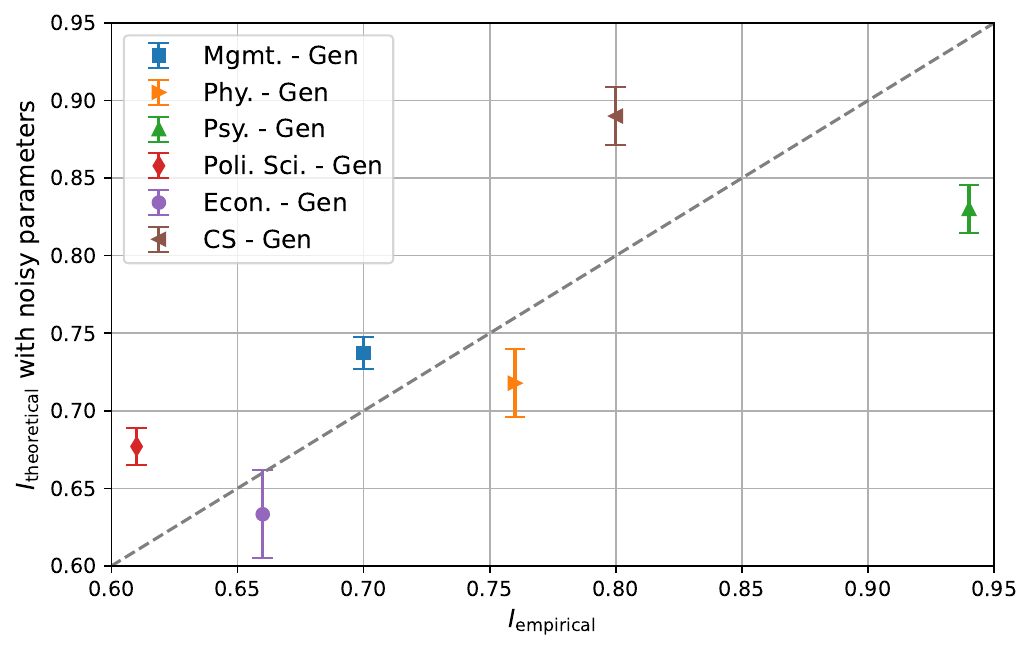}
	\caption{The power-disparity values and their confidence intervals that account for the worst-case probabilistic coverage of the gender API~(given in Table~\ref{fig:data_coverage}) and the variance introduced by the parameter estimation procedure~(described in Sec.~\ref{SI:sec:parameter_estimation}). 
	} 
	\label{fig:PowerInequalityTheoretical_CI}
\end{figure}

Recall from Figure~\ref{fig:data_coverage} that the coverage of the gender API remains above 90\% in all fields except for Computer Science. For Computer Science, coverage starts from 65\% in 1990 and then improves to 85\%. Thus, the worst-case gap in coverage of the gender API is 35\% in CS in 1990. To be conservative in our robustness tests, we assume that there is a 35\% error in the gender-API in each field~(although all fields except CS has a coverage above 90\%). More precisely, we model the gap in coverage as a normally distributed noise which causes $\redBirthProb$ to deviate in the range $\pm 0.35$ with 95\% probability. Additionally, we also recall that our empirical estimation of parameters has a standard deviation of $10^{-4}$. Again, being conservative, we randomize other parameters of the model with a normally distributed noise of standard deviation $10^{-3}$~(an order of magnitude greater than the observed standard deviation). With these noisy parameters, we compute the power-disparity using the proposed method via a Monte-Carlo simulation of 500 iterations and compute the 95\% confidence intervals of the power-disparity. The confidence intervals together with the empirically estimated values are shown in Fig.~\ref{fig:PowerInequalityTheoretical_CI}. We note that the 95\% confidence intervals calculated assuming the worst-case scenarios are not large enough to invalidate the empirical conclusions. In particular, the confidence intervals are sufficiently small~(smaller than $\pm0.03$) to illustrate a clear agreement between the empirically estimated~(in a model-agnostic manner) and theoretically predicted~(via the model-based approach) power-disparity values. This experiment illustrates the robustness of the empirical parameter estimation method to the errors introduced by the gender API and the uncertainty introduced by the empirical estimation approach.

\subsection{An Additional Sanity Check for the Empirical Results}

As a sanity check, we also compare our estimated fractions of female authors $\redBirthProb$ with the fractions of female faculty reported by independent studies. The results are summarized in the Table~\ref{tab:comparison_of_r}. It can be seen that our estimated fraction of female authors~$\redBirthProb$ is remarkably close to the fraction of female faculty independently reported in literature. We note that these cannot be expected to be exactly same since the values in literature reports the female faculty whereas we consider female authors. Additionally, the values from literature are based on data collected during different time periods. However, the fact that our estimates are aligned closely with the analogous proxy measures reported in literature~(using survey-based data collection) serve as a sanity check to instill more confidence in our empirical parameter estimation. 

\begin{table}
\centering
\captionsetup{justification=justified} 
 \begin{tabular}{l|rrrr}
 
\toprule
{} &  Estimated $\redBirthProb$ &  Literature & Source and Details\\
Field                                                           \\
\midrule
Management        &            0.35 &    0.295&                         \cite{krishen2020story}[Table~1]\\
Physics               &            0.16 &                             0.16 &        \cite{porter2019women} (based on data from 2019) & \\
Psychology       &            0.50&                             0.47 &        as of 2013; \cite{APA2014}[page~436]  \\
Political Science &            0.34 &                              0.27& as of 2015; \cite{teele2017gender}         \\
Economics         &            0.28 &                              0.224&    based on 2015 data in \cite{mcelroy2016committee}[Table~5]      \\
Computer Science               &            0.26 &                              0.224 & based on 2019 data from \cite{zippia2022}          \\
\bottomrule

\end{tabular}
\vspace{0.1cm}
\caption{\raggedright
Comparison of estimated fraction of female authors~(parameter $r$ in the model) with the fraction of female faculty members reported independently in the literature. These two values can be thought of as proxies for each other and their close agreement serves as a sanity check of our empirical estimation method utilizing the Gender API.}
\label{tab:comparison_of_r}
\end{table}

\vspace{-0.0cm}
\subsection{Reproducibility}
All codes and complete datasets are available on GitHub to facilitate reproducibility: \\\url{https://github.com/ninoch/DMPA}

\end{document}